# Cryopumping and Vacuum Systems

*Vincent Baglin*
CERN, Geneva, Switzerland

**Abstract**

The understanding of complex and/or large vacuum systems operating at cryogenic temperatures requires a specific knowledge of the vacuum science at such temperatures. At room temperature, molecules with a low binding energy to a surface are not pumped. However, at cryogenic temperatures, their sojourn time is significantly increased, thanks to the temperature reduction, which allow a "cryopumping". This lecture gives an introduction to the field of cryogenic vacuum, discussing surface desorption, sticking probability, thermal transpiration, adsorption isotherms, vapour pressure of usual gases, industrial surfaces and roughness factors. These aspects are illustrated with the case of the Large Hardon Collider explaining its beam screen and its cryosorber, leaks and beam vacuum system modelling in a cryogenic environment. Finally, operation of cryogenic beam vacuum systems is discussed for LHC and other cryogenic machines.

## 1 Introduction

This lecture is a modified, corrected, updated and extended version of a lecture given in 2006 at the CERN Accelerator School on Vacuum in Accelerators [1].

At sufficiently low temperature, the vacuum in vessels can be achieved by the adsorption of the molecules due to the attractive van der Waals forces. One of the first applications of the cryopumping mechanism was the realisation of storage containers by Sir J. Dewar after the production of the liquid hydrogen in the early 1900's [2]. At the end of the 50's, the early space projects and the requirement for producing large pumping speed in space simulating vacuum chambers at 80 K gave rise to the building of the first large cryopump cooled by helium at T < 20 K [3, 4]. In the meantime, with the invention of the Gifford-McMahon process [5] used to refrigerated low-noise amplifiers, the modern cryopump was born.

Superconductivity was discovered in 1911 by H. K. Onnes who observed at liquid helium temperature, 4.2K, the disappearance of the mercury resistivity [6]. Later on, in the late 50's, with the discovery of superconducting materials that can support large current density to produce high magnetic field, first superconducting magnets using Nb and $Nb_3Sn$ wires were built. In the meantime, first superconducting RF cavities were built to characterise the properties of superconductor materials.

Following this pioneering work in several laboratories across the world, the first superconducting RF cavities used in the context of accelerator application were built in 1974 for the Stanford superconducting accelerator, California, USA [7]. The first application of superconducting magnets was for particle detection inside a bubble chamber at Argonne, Illinois, USA [8]. Later on, the superconducting magnet technology was used for the construction of high energy storage rings. In chronological order, 1) the Tevatron at Fermilab, Illinois, USA, first beam in July 1983, 2) HERA at Desy, Hamburg, Germany, first beam in April 1991, 3) RHIC at Brookhaven, Upton, New-York, USA, first beam in June 2000 and, 4) the LHC at CERN, Geneva, Switzerland, first beam in September 2008. At present, several accelerator projects and studies, using the cryogenic technology, are ongoing across the world. The knowledge of the cryogenic-vacuum science and technology is therefore of primary importance for the vacuum scientist and engineer.

The Large Hadron Collider (LHC) [9], under operation at CERN, is one of the world's largest

instrument which rely on vacuum technology at cryogenic temperature and contributes to the development of this field. This storage ring collides two 0.58 A protons beams at 14 TeV in the centre of mass at a luminosity of $10^{34}$ Hz/cm$^2$. By means of 8.6 T superconducting dipole magnets operating at 1.9 K, the particles are kept on their 26.7 km circumference orbit. The vacuum beam lifetime, dominated by proton-nuclear scattering on the residual gas, larger than 100 h, ensures a proper operation of the machine. To fulfil this requirement, the LHC vacuum system was carefully studied and designed during more than a decade, see Refs. [10, 11]. Nowadays, thanks to a detailed understanding of the beam-gas interactions inside the cryogenic beam tube and to the excellent performance of the other systems, the LHC is routinely operating above its ultimate luminosity set at $1.8 \cdot 10^{34}$ Hz/cm$^2$, see Ref. [12]. The LHC vacuum system will therefore be extensively used to illustrate this lecture on cryogenic vacuum systems.

In the next section, some elements of cryopumping are given. The molecular desorption from a surface, the regimes of cryopumping, the sticking probability, the capture factor, and the thermal transpiration are discussed and illustrated by examples taken from the literature.

In the following section, the adsorption isotherms are described in detail. Some models of adsorption isotherms and the saturated vapour pressure are introduced. Hydrogen and helium adsorption isotherms are discussed in detail for technical surfaces used in vacuum technology. The relevance of some parameters such as the gas species, the temperature and the substrate porosity are underlined using literature's data.

The last section discusses different aspects of some cryo-vacuum systems. The LHC specificities owing to the different cold bore operating temperature at 1.9 K and 4.5 K are described and explained. Helium leaks in cryogenically cooled tube are described and illustrated together with their impact on a vacuum system design. Design and modelling of cryogenic beam tubes is explained and illustrated using LHC. Finally, a few examples on the specificities of the cryo-vacuum operation for some machines are discussed.

Looking at Fig. 1, that shows a white stain at an extremity of the LHC beam screen, if you wonder:

- What is this white stain?
- Why it is on the LHC beam screen?
- What is then the expected gas density in this expensive vacuum system?
- How to avoid the growth of this stain?
- How to get rid of it?
- What will happen when the beam will circulate?

This lecture may help you to find it out!

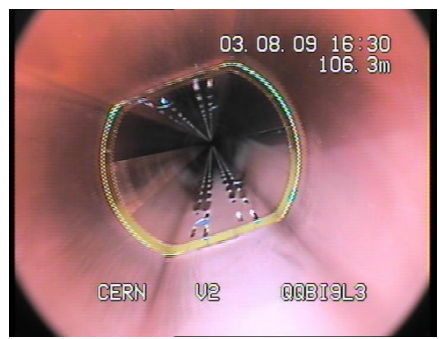

**Fig. 1**: White stain on the LHC beam screen following an "exotic" intervention for repair.

## 2 Elements of cryopumping

### 2.1 Desorption from a surface

Molecules can be bounded to a surface. When a molecule is desorbed from a surface at a given activation energy of desorption (or binding energy), E, its rate of desorption is described by the Polanyi-Wigner equation

$$\frac{d\theta}{dt} = -\theta \, \nu_0 \, e^{-\frac{E}{kT}} \tag{1}$$

where $\nu_0$ is the frequency of vibration of the molecule, k the Boltzmann constant (86.17 $10^{-6}$ eV/K) and T the temperature of the surface. The frequency $\nu_0$ is expected to be ~ $10^{13}$ Hz at room temperature but can increase to ~ $10^{16}$ Hz for a localised adsorption of a rather immobile molecule. The equation is given for a first-order desorption in which the desorption rate depends linearly on the surface coverage $\theta$. The first-order desorption applies for physisorbed molecules and for non-dissociated chemisorbed molecules ($H_2$, $N_2$ and $O_2$ are dissociatively chemisorbed on metals). Second order kinetics accounts for recombinative desorption of chemisorbed atoms and is outside the scope of this lecture, see Ref. [13].

The desorption process is characterised by the sojourn time, $\tau$ given by the Frenkel Eq. (2). This relation shows that the higher the temperature, the shorter the sojourn time is. This fact is routinely observed in vacuum technology during a bake out when mainly water molecules are desorbed from the surface above 100°C. Moreover, the equation shows that molecules with low binding energy *i.e.* physisorbed molecules, have large sojourn time at low temperature, explaining the cryopumping effect.

$$\tau = \frac{e^{\frac{E}{kT}}}{\nu_0} \tag{2}$$

Table 1 gives some activation energies of desorption of physisorbed (at low surface coverage) and chemisorbed molecules on technical surfaces. The physisorbed molecules have activation energy of desorption in the range of ~0.1 eV and the chemisorbed molecules are in the range of ~1 eV. As shown by Redhead, the activation energy can be measured by Temperature Programmed Desorption (TPD) or Thermal Desorption Spectroscopy (TDS) [14]. The process consists in recording the pressure while the sample's temperature, T, is increased at a constant rate, $\beta$, *i.e.* $T = T_0 + \beta t$. At the maximum of the desorption rate *i.e.* at the pressure peak of the desorption spectrum, the temperature, $T_P$, is recorded and the activation energy of the desorption is computed using Eq. (3). This equation is obtained from the derivative of Eq. (1) which equals zero at the maximum desorption rate.

$$\frac{E}{kT_P^2} = \frac{\nu_0}{\beta} e^{-\frac{E}{kT_P}} \tag{3}$$

As a rule of thumb, an order of magnitude of the activation energy, in meV, can be obtained from Eq. (4) with $T_P$ in K.

$$E \sim 2.6 \, T_P \tag{4}$$

In the case of physisorbed molecules, a large sojourn time, above 100 h, is obtained for a temperature of the substrate in the range of 5 to 65 K. Hence, the pumping of most of the gases by physisorption is achieved below 65 K. A simple cooling of a metallic surface to the liquid nitrogen temperature (77 K) will only pump water molecules. Physisorption occurs below 20 K for binding energies < 0.1 eV, below 50 K for binding energy < 0.2 eV and below 70 K for molecules with binding energy < 0.3 eV.

In an unbaked vacuum system, the residual gas composition is dominated by water which is chemisorbed on the vacuum chamber walls held at room temperature. At this temperature, the sojourn time is 100 h for water while it is only 3 minutes for hydrogen and more than 6 months for carbon monoxide. Increasing the temperature of the vacuum vessel above 100 ºC reduces the sojourn time of water to a minute, hence, strongly reduces its surface coverage and thus the outgassing rate of water at room temperature after bake-out. However, molecules which are strongly bounded to the material's surface can be desorbed by stimulated desorption (in an accelerator for instance). Therefore, a bake out up to at least 300 ºC is required to reduce the stimulated desorption of carbon monoxide molecules with 1.7 eV binding energy.

**Table 1:** Activation energies of desorption, E, of some physisorbed (with low surface coverage) and chemisorbed molecules.

|              |        | E [eV] | Surface              | Ref      |
|--------------|--------|--------|----------------------|----------|
| Physisorbed  | $H_2$  | 0.017  | Stainless steel 316 L| [15]     |
|              | $H_2$  | 0.065  | Al anodised          | [15]     |
|              | $H_2$  | 0.081  | Saran Charcoal       | [13]     |
|              | $H_2$  | 0.137  | Amorphous carbon     | [16]     |
|              | $H_2$  | 0.185  | Activated charcoal   | [17]     |
|              | $H_2$  | 0.23   | Carbon fibre         | [18]     |
|              | CO     | 0.365  | Amorphous carbon     | [16]     |
| Chemisorbed  | $H_2$  | 0.9    | Stainless steel 304 L| [19, 20] |
|              | $H_2O$ | 1.1    | Aluminium            | [21]     |
|              | CO     | 1.2    | Stainless steel 316 L| [22]     |
|              | CO     | 1.7    | Stainless steel 316 L| [22]     |

Fig. 2 shows the desorbed gas following a natural warm-up from 10 K to room temperature of a stainless-steel pipe. In this case, the unbaked system was evacuated to $5\times10^{-9}$ mbar, cool down to 3 K for about 3 days and then let warming up at a rate of ~ 2.2 K/h. Hydrogen is the first gas to desorb followed closely by carbon monoxide and methane. Later comes the carbon dioxide desorption peak and much later the water desorption. The number of desorbed molecules range from $5\times10^{12}$ to $2\times10^{14}$ molecules/cm$^2$.

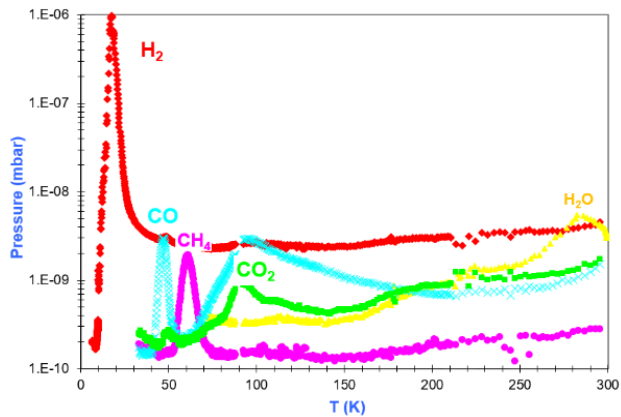

**Fig. 2**: Molecular species desorbed during a natural warm-up from 10 K to room temperature.

Table 2 gives the peak temperature and the corresponding activation energy of the desorbed molecular species shown in Fig. 1. The respective boiling points and evaporation heat at boiling point of these molecules are also given for comparison. Molecules can be cryopumped on cold surfaces with different binding energies.

Table 2: Some properties of molecules when adsorbed at low surface coverage and liquid phase.

|  | Molecules | $H_2$ | $CH_4$ | $H_2O$ | CO | $CO_2$ |
|---|---|---|---|---|---|---|
| Low surface coverage | Peak temperature [K] | 18 | 60 | 285 | 45 | 95 |
|  | Activation energy [meV] | 57 | 199 | 965 | 152 | 303 |
| Liquid | Boiling point [K] | 22.3 | 111.7 | 373.2 | 81.7 | 195.1 |
|  | Evaporation heat at boiling point [meV] | 9.3 | 89.4 | 421.6 | 62.7 | 158.9 |

## 2.2 Cryopumping regimes

At cryogenic temperature, a molecule interacting with a surface is physisorbed, due to the Van der Walls force, if the surface is "cold enough". Several regimes of cryopumping are illustrated in Fig. 3 and defined below, physisorption, condensation and cryotrapping.

- Physisorption is the regime of the sub-monolayer coverage. The van der Waals force acts between the adsorbed molecule and the material surface. The binding energies are those given in Table 1. For hydrogen, the binding energy varies from 20 to 85 meV for smooth and porous material. One-hour sojourn time is obtained at 5 K and 26 K respectively. Since the binding energy for physisorption is larger than the heat of vaporisation (9.3 meV for $H_2$), sub-monolayer quantities of all gas can be physisorbed in sub-saturated conditions at their boiling temperature. Pumps relying on physisorption are called cryosorption pumps [23].

- When the surface coverage increases, the van der Walls force acts between the molecules themselves. This is the cryocondensation regime. While increasing the surface coverage, a saturation equilibrium between gas adsorption and desorption is reached. This is the saturated vapour pressure. The activation energy of desorption equals the energy of vaporisation. It ranges from 10 to 160 meV for hydrogen and carbon dioxide respectively. Cryocondensation is the pumping of gas due to a phase change into a solid or a liquid (*i.e.* an ice or frost of the gas). Pumps relying on condensation are called cryocondensation pumps [24].

- The possibility to use a condensable gas to trap a non-condensable gas with a high vapour pressure is called cryotrapping. For instance, argon is used to trap helium and hydrogen in cryopumps. The trapped molecules are incorporated in the condensable carrier so that the equilibrium pressure is significantly lower than in pure physisorption.

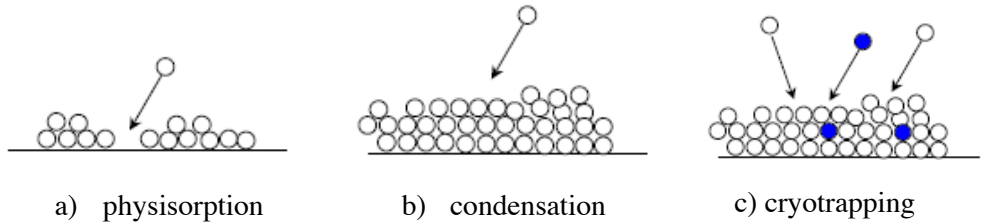

a) physisorption    b) condensation    c) cryotrapping

**Fig. 3**: Illustration of the different cryopumping regimes

## 2.3 Sticking probability, pumping speed and capture factor

The sticking probability (or sticking coefficient) for cryosorption and cryocondensation is the ratio of the average number of molecules which stick on a cold surface divided by the total number of impinging molecules. The sticking probability shall not be confused with the vapour pressure which is another physical phenomena. It varies between 0 and 1. The value is a function of the gas species, the surface nature, the surface coverage, the temperature of the gas and the surface temperature. Fig. 4 shows the sticking probability of hydrogen at 300 K incident onto a surface at 3.1 K as a function of surface coverage, see Ref. [25]. The sticking probability increases with the surface coverage. For thick surface coverage above $10^{17}$ H$_2$/cm$^2$ (~ 100 monolayers), the sticking probability approaches unity. It is then named condensation coefficient since the pumping is in the cryocondensation regime. Fig. 5 shows the variation of the hydrogen condensation coefficient onto a surface at ~ 4 K for different temperatures (*i.e.* energies or velocities) of the incoming hydrogen, see Ref. [26]. For low energies, the condensation coefficient is close to one. It is still above 0.8 for hydrogen held at room temperature.

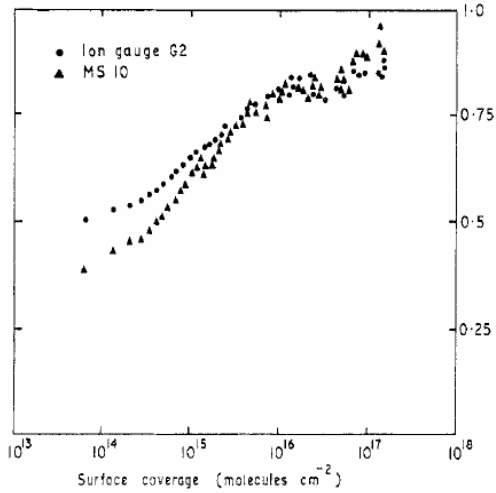 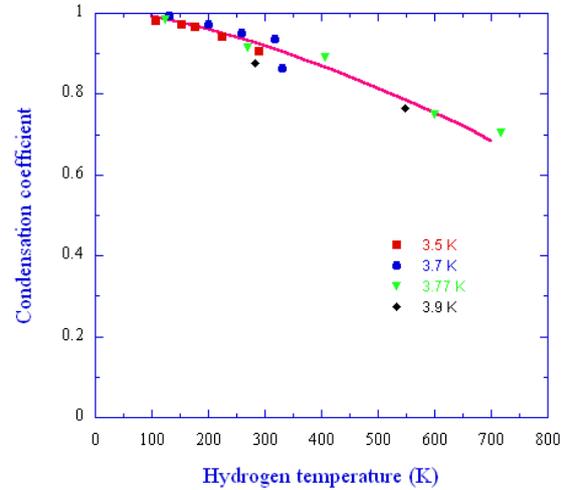

**Fig. 4**: Sticking probability of hydrogen as a function of the surface coverage. The hydrogen is at 300 K incident onto a surface at 3.1 K [25].

**Fig. 5**: Hydrogen condensation coefficient onto a surface at ~ 4 K as a function of the hydrogen temperature [26].

The availability of sticking coefficient data in the literature is rather limited, in particular below a monolayer. Apart from the values given in this lecture, the interested reader may find other data in the following Refs. [27, 28, 29, 30, 31].

Since a cold surface acts as a pump, its pumping speed can be evaluated from the gas kinetics theory. If the temperature at the surface of the cryosorbed gas is so high such that its vapour pressure, $P_{vap}$, becomes comparable with the pressure of the incident gas, P, the effective pumping speed, S, of the cryopump is given by Eq. (4) and reduces to zero when $P=P_{vap}$. However, most of the cryopumps does not operate in such conditions so that the pumping speed is simplified to the product of the sticking probability, σ, times the aperture conductance of the cold surface.

$$S = \frac{1}{4} \sigma \left(1 - \frac{P}{P_{vap}}\right) A \overline{V} \approx \frac{1}{4} \sigma A \overline{V} \quad (4)$$

where A is the geometrical area of the surface and $\overline{V}$ the mean molecular velocity. A practical formula of the pumping speed in ℓ.s$^{-1}$.cm$^{-2}$ is given by

$$S = 3.64 \, \sigma \sqrt{\frac{T}{M}} \quad (5)$$

where T is the temperature of the surface (*i.e.* when the temperature of the gas is accommodated to the surface temperature) and M the molecular weight of the incident gas. At 4.2 K, the maximum pumping speed of hydrogen and carbon monoxide equals 5.3 and 1.4 ℓ.s$^{-1}$.cm$^{-2}$.

In a cryopump, to minimise the heat load, the cold surface is shielded from the part of the vacuum vessel held at room temperature. Therefore, the molecules shall find their way towards the cold surface to be pumped. This pure geometrical effect is taken into account by the so-called capture factor (or capture probability), $C_f$, which takes into account the conductance, C, of the system as shown in Eq. (6).

$$C_f = \frac{C\,\sigma}{C+\sigma} \tag{6}$$

The computing of the capture probability is required to optimise the design of the vacuum system. Monte-Carlo or angular coefficients methods can be used to compute the capture probability of complex systems, see Ref. [32]. For instance, a cryopump head is shielded from the room temperature radiation heat load by chevron type baffles. In this case, the pumping speed of the cryopump is reduced compared to the pumping speed of the cold surface. The capture probability of such baffles is estimated to be 0.3, see Refs. [33, 34]. Another example is the case of the LHC beam screen inserted into its cold bore, here, the angular coefficients method was used to optimise the position of the holes in the electron shield with respect to the position of the pumping holes in the LHC beam screen, see Ref. [35].

## 2.4 Thermal transpiration

To reduce the heat load on the cryogenic system, the monitoring of a cryogenic vacuum system is usually made by vacuum gauges located in a room temperature vacuum vessel. In this case, a thermal transpiration correction shall be applied (the so-called Knudsen relationship), see Ref. [36]. When two vessels at two different temperatures $T_1$ and $T_2$ are connected by a small aperture, the collision rate at the aperture is conserved when the steady state is established. By equating the fluxes, and since the mean velocity scales like $\sqrt{T}$, the ratio of the pressure $P_1/P_2$ and the gas density $n_1/n_2$ in the two vessels are given by

$$\frac{P_1}{P_2} = \sqrt{\frac{T_1}{T_2}} \text{ and } \frac{n_1}{n_2} = \sqrt{\frac{T_2}{T_1}} \tag{7}$$

In practice, when measured at room temperature, the pressure inside a vacuum vessel cooled with liquid helium (4.2 K) is the measured pressure divided by 8. The measured pressure shall be divided by 2 for a vacuum vessel cooled with liquid nitrogen (77 K).

Fig. 6 illustrates the validity of the thermal transpiration correction in the UHV-XHV range. The figure shows the adsorption isotherms of hydrogen at 4.2 K measured by a gauge located at room temperature and a gauge immersed in liquid helium at 4.2 K, see Ref. [37]. For such isotherms measurements, known quantities of hydrogen are successively admitted into the cold system and the equilibrium pressure (or vapour pressure) measured by both gauges. For ease of comparison, the thermal transpiration correction is applied to the gauge located at room temperature. The pressure is therefore given in both cases at a temperature of 4.2 K. As expected from the thermal transpiration theory, the values are similar for the two gauges readings over, at least, 6 decades from $10^{-12}$ until $10^{-6}$ Torr! The lowest pressure, $6 \cdot 10^{-14}$ Torr at 4.2 K (equivalent to $5 \cdot 10^{-13}$ Torr at room temperature), was measured by the extractor gauge held at 4.2 K.

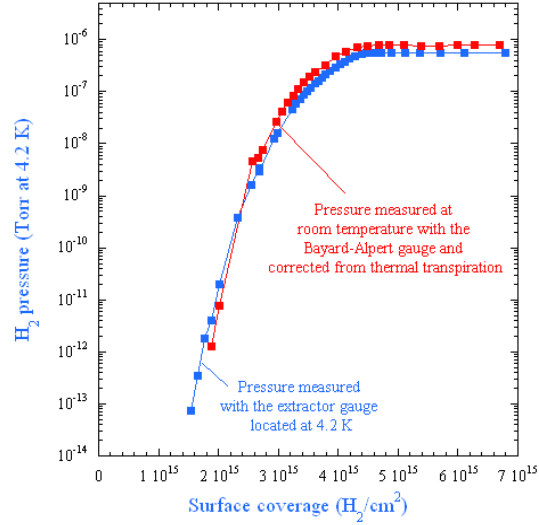

**Fig. 6**: Adsorption isotherms of $H_2$ at 4.2 K measured by an ionisation gauge located at room temperature and an ionisation gauge immersed in liquid helium at 4.2 K [37].

## 3 Adsorption isotherms
### 3.1 Some models

The measurement at constant temperature of the equilibrium pressure (vapour pressure) as a function of the surface coverage is an adsorption isotherm. The adsorption isotherm is a function of the molecular species, the temperature of the surface, the nature of the surface and the gas composition. In the following, some measurements of the isotherms are extracted from the literature to illustrate the effects of the various parameters. The examples are limited to the hydrogen and the helium since they are the main gases present in a vacuum system held at liquid helium temperature.

There are several semi-empirical adsorption isotherms models. At low surface coverage, when there are no lateral interactions between the adsorbed gas molecules, the vapour pressure follows the Henry's law. It predicts that the surface coverage, $\theta$, will vary linearly with the pressure, P.

$$\theta = c\,P \tag{8}$$

In the sub-monolayer range (illustration a) in Fig. 3), for metallic, glass and porous substrate, the isotherm is better described by the DRK (Dubinin, Raduskevic and Kanager) model. The model is valid at low pressure and offers a good prediction of the isotherm as a function of the temperature [38]. The DRK Eq. (9) is a function of the surface coverage, $\theta$, the monolayer surface coverage, $\theta_m$, the surface temperature, T, the saturated vapour pressure, $P_{sat}$, and the pressure, P, and a constant D. A plot of the measured isotherm in the DRK coordinates *i.e.* $\ln(\theta)$ vs. $\varepsilon^2$, yields a straight line from which the constant D and the monolayer capacity can be computed.

$$\ln(\theta) = \ln(\theta_m) - D\varepsilon^2 = \ln(\theta_m) - D\left(kT\ln\left(\frac{P_{sat}}{P}\right)\right)^2 \tag{9}$$

The BET (Brunauer, Emmet and Teller) is a multimonolayer description of the isotherm (illustration b) in Fig. 3). It is a function of the surface coverage, $\theta$, the monolayer surface coverage, $\theta_m$, the saturated vapour pressure, $P_{sat}$, and the pressure, P, and a constant $\alpha$. The constant $\alpha$ is much larger than one such that the BET Eq. (10) can be simplified. In this case also, the monolayer capacity can be computed from a linear plot in the BET coordinates ($P/(\theta(P_{sat}-P))$ vs. $P/P_{sat}$). It shall be noted that the monolayer surface coverage derived from the BET model is different from the one derived by the DRK model. Despite the BET model covers the entire pressure range from Henry's law to the saturated vapour pressure of the adsorbate, it is only an adequate description of the experimental data in the range $0.01 < P/P_{sat} < 0.3$ [39].

$$\frac{P}{\theta(P_{sat}-P)} = \frac{1}{\alpha\,\theta_m} + \frac{(\alpha-1)}{\alpha\,\theta_m}\frac{P}{P_{sat}} \approx \frac{1}{\theta_m}\frac{P}{P_{sat}} \qquad (10)$$

## 3.2 Saturated vapour pressure

The saturated vapour pressure of a gas is the pressure of this gas over its liquid or its solid phase *i.e.* when many monolayers of gas are condensed on a substrate. The saturated vapour pressure follows the Clausius-Clapeyron equation where A and B are constants and T the temperature of the condensing surface.

$$\mathrm{Log}(P_{sat}) = A - \frac{B}{T} \qquad (11)$$

Fig. 7 shows the saturated vapour pressure of the some common gases as a function of the surface temperature [13]. In most cases, the data were obtained for metallic (e.g Cu) or glass (e.g Pyrex) surfaces and measured in the temperature range corresponding to a saturated vapour pressure above $10^{-5}$-$10^{-4}$ Torr except for He and $H_2$ where data were measured below their evaporation temperature at atmospheric pressure [40, 41]. The curve shown here in the XHV-UHV range are extrapolation and therefore indicative. The pressures, corrected from the thermal transpiration, are given at room temperature and expressed in Torr. The figure indicates that below 20 K, the saturated vapour pressure of most of the gases is below $10^{-13}$ Torr. Therefore, large quantities of such gases can be pumped below 20 K. At liquid helium temperature (4.2 K) only helium and hydrogen are not pumped. At 4.2 K, the saturated vapour pressure of hydrogen is $\sim 10^{-6}$ Torr as already shown in Fig. 6.

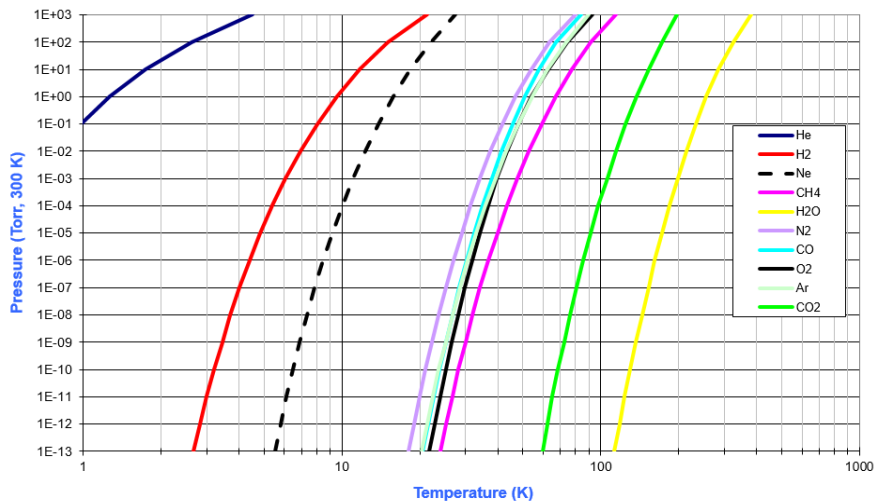

**Fig. 7**: Saturated vapour pressure curves as a function of the temperature [13, 40, 41].

Table 3 gives the A and B constants (derived from Refs. [13, 40], for He and [41] for $H_2$) to be use, by the vacuum engineer, in Eq. (11) to compute the saturated vapour pressure (expressed in mbar at 300 K) for some common gas condensed on a vacuum surface.

**Table 3**: A and B constants of some usual molecular species when the saturated vapour pressure is expressed in mbar at 300 K.

|   | He | $H_2$ | $CH_4$ | $H_2O$ | Ne | $N_2$ | CO | $C_2H_6$ | $O_2$ | Ar | $CO_2$ |
|---|----|----|----|----|----|----|----|----|----|----|----|
| A | 4.09 | 3.97 | 7.36 | 10.23 | 6.95 | 8.14 | 8.67 | 9.69 | 8.57 | 7.84 | 9.995 |
| B | 4.96 | 40.9 | 486.80 | 2612.60 | 109.87 | 379.27 | 441.19 | 1039.34 | 463.448 | 420.58 | 1360.35 |

Two other observations, that will be later discussed in more detail, may be done from the previous figure:

- In superconducting machines, the saturated vapour of hydrogen may span over 12 decades from $1 \; 10^{-17}$ mbar at 1.9 K to $4 \; 10^{-5}$ mbar at 4.5 K. Below 2.7 K, the hydrogen vapour pressure falls in the XHV range ($10^{-12}$ mbar) and is therefore negligible for most accelerator application. Above this operating temperature, the condensed hydrogen may have some impact on the vacuum system design as discussed in Sections 4.1 and 4.2.
- Considering helium at 1.9 K, the saturated vapour pressure is in the low vacuum range (23 mbar) and therefore He leaks into any cryogenic vacuum system are unacceptable (see Section 4.3).

### 3.3 Hydrogen adsorption isotherms

As shown in Fig. 7, by pumping on a liquid helium bath, the temperature of the liquid helium can be reduced to allow the pumping of large quantities of hydrogen with a condensation cryopump, see Ref. [42]. A condensation cryopump which operated from 2.3 K (50 Torr on the helium bath) to 4.2 K (helium atmospheric pressure) was used to measure the adsorption isotherm of hydrogen as a function of temperature, see Ref. [41].

Fig. 8 shows the adsorption isotherm of hydrogen onto a stainless-steel surface as a function of the temperature. The monolayer capacity is estimated to be $3 \; 10^{15}$ $H_2/cm^2$. Above $10^{16}$ $H_2/cm^2$ the saturated vapour pressure is reached. Below 3 K, the saturated vapour pressure of hydrogen is negligible. This saturated vapour pressure follows the Clausius-Clapeyron law illustrated in Fig. 7 and expand over five decades. At 4.17 K, the saturated vapour pressure of hydrogen equals $7 \; 10^{-7}$ Torr ($9 \; 10^{-7}$ mbar) which translates into $6 \; 10^{-6}$ Torr ($8 \; 10^{-6}$ mbar) when thermal transpiration corrections at 300 K are applied. The gas density is the quantity of interest for a vacuum designer dealing with cryogenic beam vacuum systems; the saturated gas density of a surface held at 4.17 K equals $1.6 \; 10^{18}$ $H_2/m^3$.

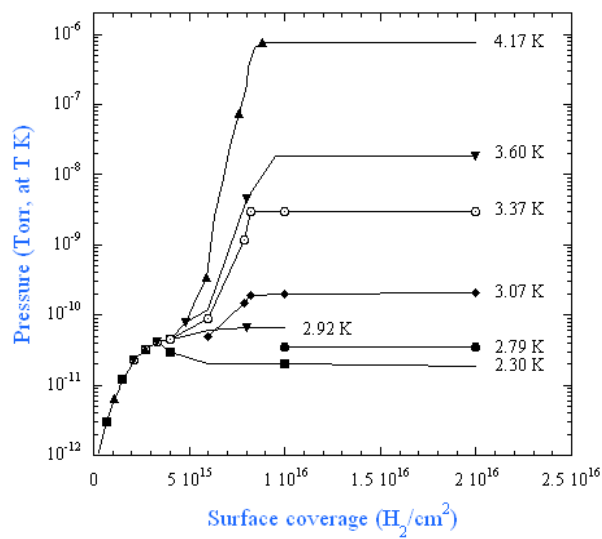

**Fig. 8**: Hydrogen adsorption isotherm onto a stainless-steel surface as a function of the temperature [41].

The addition of a gas can drastically modifies the shape of the hydrogen isotherm [43]. Detailed studies using a real machine environment are therefore mandatory in order to predict the behaviour of the designed cryogenic system.

For example, a frost of carbon dioxide, which is condensed onto the surface prior to the hydrogen injection, forms a porous layer. As shown in Fig. 9, the adsorption capacity of the electroplated Cu surface is increased by the addition of a carbon dioxide frost. The steep pressure rise is indeed shifted from $2\ 10^{15}$ to $6\ 10^{15}$ $H_2/cm_2$. The estimated DRK adsorption capacity of the condensate is 0.3 $H_2/CO_2$. Ar frost is routinely used in cryo-vacuum technology of large instruments [44].

Similarly, the admission of gas in the vacuum chamber might change the shape of the adsorption isotherm. This is illustrated again with the case of carbon dioxide mixed with hydrogen. As shown in Fig. 10, injecting a mixture of 45 % $H_2$ and 55 % $CO_2$ decrease the level of the saturated vapour pressure by 2 orders of magnitude. This phenomenon is applied in cryopumps when carbon dioxide is injected to enhance the pumping of hydrogen and helium by cryotrapping.

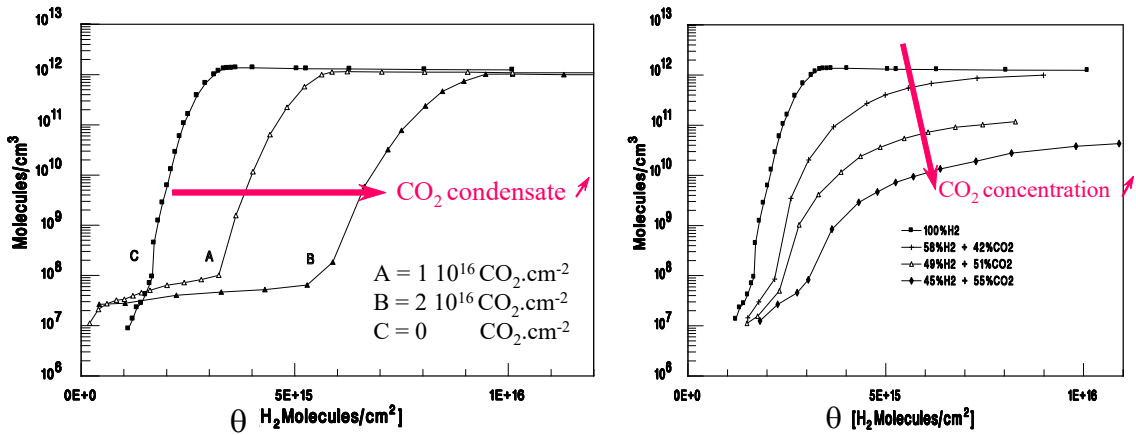

**Fig. 9**: $H_2$ adsorption isotherm at 4.2 K on a pre-condensed layer of $CO_2$ [43].

**Fig. 10**: $H_2$ co-adsorption isotherm at 4.2 K with $CO_2$ [43].

Recently, for the construction at Darmstadt, Germany, of the superconducting heavy ion synchrotron SIS 100 at Fair, $H_2$ adsorption isotherm on electropolished stainless steel have been obtained in the 7 to 18 K range on a stainless-steel surface [45].

At these temperatures, a vapour pressure as large as $10^{-7}$-$10^{-6}$ mbar is reached in the sub-monolayer range ($10^{13}$ - $4\ 10^{14}$ $H_2/cm^2$). As expected, the measured isotherms are well described by a DRK model with $\theta_m = 6.5\ 10^{14}$ $H_2/cm^2$ and $D = 3075$ $eV^{-2}$ in rather good agreement with previous theoretical prediction for a heterogeneous surface with $D = 1907.5$ $eV^{-2}$ or $D^{-1/2} = 528$ cal/mol, see Ref. [46].

## 3.4 Helium adsorption isotherm

Another gas of interest at cryogenic temperature is helium. As depicted by the saturated vapour pressure of helium in Fig. 7, one shall operate below 4 K to reach a low vapour pressure of helium. Thus, in order to perform measurements in UHV condition, one must work in the sub-monolayer range providing in the meantime, a good test for the validity of the Henry's law.

Fig. 11 shows the helium adsorption isotherm on a stainless-steel tube as a function of temperature [47]. Similar to hydrogen, vapour pressure in the High-Vacuum regime is reached for sub-monolayer adsorption. In this case also, the isotherms are well described by the DRK model with $\theta_m = 1.3 \cdot 10^{15}$ He/cm$^2$ and $D = 2.92 \cdot 10^4$ eV$^{-2}$. Below $10^{-9}$ Torr and $10^{14}$ atoms/cm$^2$, the behaviour of the isotherms approaches the Henry's law with a slope of unity, see Refs. [39, 38, 46].

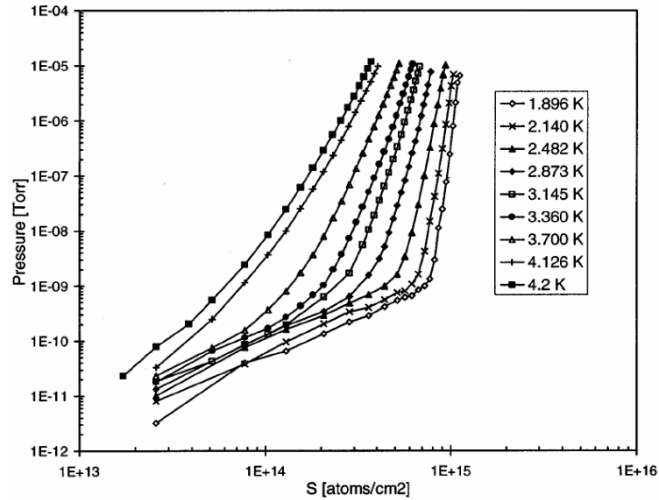

**Fig. 11**: Helium adsorption isotherm measured on a stainless-steel tube from 1.9 to 4.2 K [47].

### 3.5 Adsorption isotherms of other gases

Future superconducting circular colliders are planned to operate at high temperature (e.g. around 60 K) with the objective to reduce the cost of the cryogenic cooling and increase the molecular pumping speed, see Ref. [48]. When increasing the operating temperature of a cryogenic beam pipe, adsorption isotherms of gases other than helium and hydrogen are therefore of great interest. Unfortunately, experimental data obtained in the temperature range on metallic technical surface are still limited.

Adsorption isotherm of carbon dioxide was studied for stainless steel held at 77 K, see Ref. [30]. In the sub-monolayer region ($< 6 \cdot 10^{14}$ CO$_2$/cm$^2$), it was observed that the necessary time to reach the equilibrium pressure was much larger than for hydrogen in such a way that the equilibrium vapour pressure was obtained after more than 12h. However, above a monolayer, the saturated vapour pressure was reached almost immediately after the gas injection. The sticking coefficient measurement exhibited a linear trend with surface coverage with very low values from 0.003 at $10^{14}$ CO$_2$/cm$^2$ till 0.2 at $10^{16}$ CO$_2$/cm$^2$. The duration needed to reach equilibrium in the sub-monolayer regime together with the huge quantity of molecules required to reach a large sticking coefficient suggest a non-uniform adsorption of CO$_2$ which aggregates in multi-layers clusters of increasing size.

### 3.6 Industrial surfaces

The molecules interaction may differ significantly with the type and the nature of a surface. In particular, the nature of industrial or technical surfaces used for construction in vacuum technology can be strongly different. In this context, hydrogen adsorption isotherms and Temperature Programmed Desorption have been used to study the porosity of technical surfaces [15, 49].

The studied materials of potential use in accelerator vacuum system may be classified in smooth and porous surfaces. On one hand, electrochemical buffed 316L stainless steel, 100 $\mu$m thick copper colaminated on stainless steel and 1 $\mu$m thick TiZrV film are smooth surfaces, on the other hand, 40 $\mu$m thick anodised aluminium and 400 $\mu$m thick amorphous carbon coating are porous surfaces. The monolayer capacity of smooth surfaces is in the range $10^{15}$ H$_2$/cm$^2$ whereas the monolayer capacity of porous surface is in the range $10^{17}$-$10^{18}$ H$_2$/cm$^2$.

The difference in adsorption capacity between the materials originates from different adsorption sites. TDS studies show that the binding energy varies with the nature of the substrate, its surface coverage and its temperature. At 4.2 K, all saturated surfaces have a binding energy in the range 17-24 meV (comparable with the evaporation heat of $H_2$ at 20 K, see Table 2) but porous surfaces have others adsorption sites at a much higher energy between 50 and 100 meV. Sub-saturated porous surfaces can adsorb molecules at higher energy. Thick amorphous carbon coating can adsorb up to 5 $10^{14}$ $H_2/cm^2$ until 60 K. It is observed that by increasing the quantity of adsorbed hydrogen on a porous surface, the number and the energy level of the available adsorption sites decrease. Therefore, when increasing the surface coverage on a surface, the temperature of molecular desorption decreases as observed, for instance, with porous amorphous solid water, see Ref. [50].

### 3.7 Roughness factor

Apart from hydrogen condensation at low temperature, the hydrogen and helium molecular capacities on technical metallic surface are very small *i.e.* $10^{14}$ to $10^{15}$ molecules/cm$^2$. Cryosorbing or porous materials are therefore used in vacuum pumps to increase the capacity, the pumping speed and the operating temperature range. For instance: activated charcoal is used in cryosorption pumps. Its capacity is ~ $10^{22}$ $H_2$/g *i.e.* $10^{21}$ monolayers, see Ref. [13]. Its sticking probability equals 0.3 at 30 K, see Ref. [51]. Thanks to these material properties, activated charcoal is mounted onto cryopanels installed on the second stage of a cryopump to allow the pumping of hydrogen and helium at 20 K, see Ref. [23].

The capacity of cryopanels or technical surfaces is usually measured with the BET method. From the xenon isotherm at 77 K, the BET monolayer, $\theta_{m\text{-}BET}$, is obtained using Eq. (10). Fig. 12 shows the xenon adsorption isotherm obtained at 77 K for some technical surface. At 77 K, the saturated vapour pressure measured using a spinning rotor gauge (an absolute pressure gauge) equals 1.5 $10^{-3}$ Torr. Two categories of materials are seen, the one with low monolayer capacity (Cu, 304L, Al) and the one with large monolayer capacities (30 $\mu$m thick anodised Al, NEG).

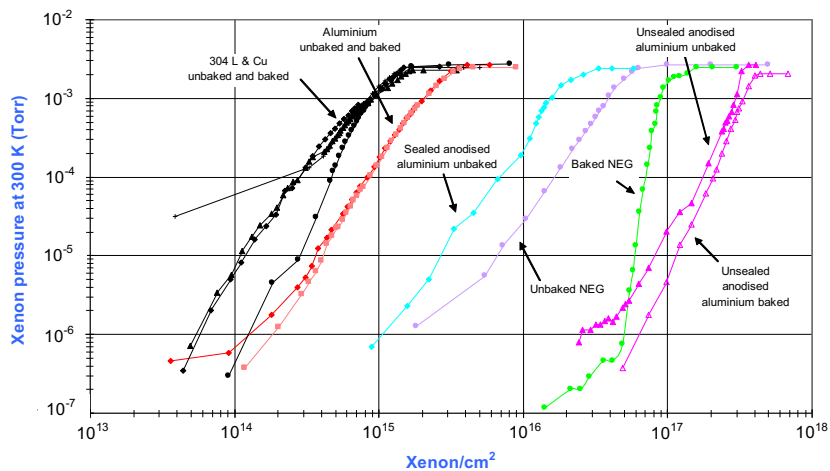

**Fig. 12:** Xenon adsorption isotherm of some technical surfaces at 77 K [52].

Table 4 gives the BET monolayer capacity computed from the slope of the above isotherms data expressed in the BET coordinates (P/P$_{sat}$, P/($\theta$(P$_{sat}$-P))).

Table 4: BET monolayer capacity for some technical surfaces

|  | Copper DHP acid etched | Stainless steel 304L vacuum | Aluminium degreased | Anodised Al sealed | Anodised Al unsealed | NEG St 707 |
|---|---|---|---|---|---|---|
| Unbaked | 5.6 $10^{14}$ | 5.3 $10^{14}$ | 1.4 $10^{15}$ | 1.0 $10^{16}$ | 2.2 $10^{17}$ | 2.8 $10^{16}$ |
| Baked at 150ºC | 7.6 $10^{14}$ | 6.0 $10^{14}$ | 1.4 $10^{15}$ | - | 2.2 $10^{17}$ | 6.3 $10^{16}$ |

Since the area of a xenon molecules, $a_{Xe}$, is about 25 Å$^2$, the roughness factor, $r_f$, i.e. the ratio of the real surface area seen by a molecule over the geometrical surface can be computed from the BET monolayer capacity ($r_f = \theta_{m\text{-}BET} \times a_{Xe}$). Table 5 shows the roughness factor of several technical surfaces used in vacuum technology. Metallic surfaces have roughness factors close to unity *i.e.* a low surface capacity whereas porous surfaces such as anodised aluminium and Non-Evaporable Getters (NEG) St 707 have large roughness factors, see Ref. [52].

Table 5: Roughness factor of some technical surfaces

|  | Copper DHP acid etched | Stainless steel 304L vacuum | Aluminium degreased | Anodised Al sealed | Anodised Al unsealed | NEG St 707 |
|---|---|---|---|---|---|---|
| Unbaked | 1.4 | 1.3 | 3.5 | 24.9 | 537.5 | 70.3 |
| Baked at 150ºC | 1.9 | 1.5 (at 300 ºC) | 3.5 | - | 556 | 156.3 |

## 3.8 Carbon fiber cryosorber

Commercially available cryopumps are made of activated coconut charcoal glued on the cryopanels. In some circumstances and for particular reasons, this product cannot be used, and other material must be selected. This is the case of the LHC vacuum system for which a dedicated study campaign was launched to select an appropriate cryosorber, see Refs. [17, 53, 54]. The selected material is a woven carbon fibre that ease installation, sustain high radiation and meet the LHC vacuum requirements, see Ref. [55].

The LHC cryosorber has the appearance of a fabric with woven carbon fibre. A carbon fibre wire is ~ 1 mm diameter. The wires are weaved. The fibres, which composed the wire, are ~ 10 μm diameter. Each fibre has pores which diameter range from 50 to 500 nm. The performance of the cryosorber being a function of the pore size distribution, the pumping speed and the capacity are provided by the trapping of the molecules within these pores, see Refs. [15, 23, 56]. Fig. 13 shows the hydrogen adsorption isotherms on a woven carbon fibre in the 6-30 K range, see Ref. [18]. The capacity, which is here defined by the amount of gas which can be pumped at an equilibrium pressure not exceeding the LHC operating pressure, range from $10^{18}$ to $10^{17}$ H$_2$/cm$^2$ at 6 and 30 K respectively. The roughness factor of such a cryosorber is about one thousand times larger than a metallic surface. Fig. 14 shows the sticking probability of the woven carbon fibre. At $10^{18}$ H$_2$/cm$^2$, the sticking probability decreases from 0.3 to 0.1 at 6 and 30 K respectively. These values are comparable with those of the activated charcoal.

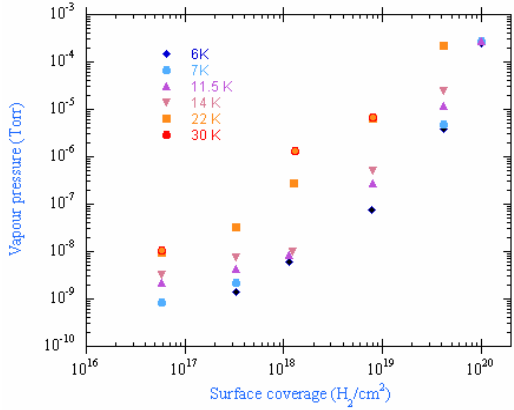 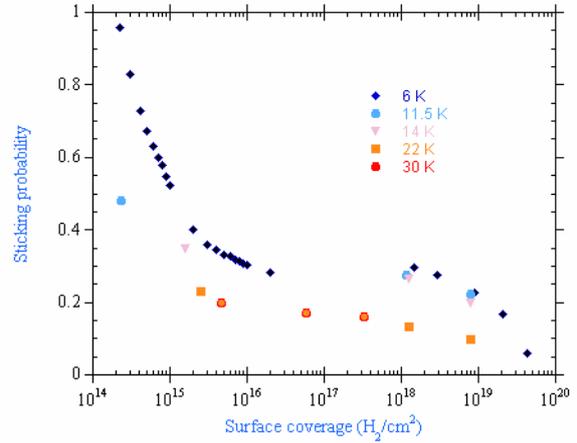

**Fig. 13**: Hydrogen adsorption isotherms on a woven carbon fiber from 6 to 30 K [18].

**Fig. 14**: Sticking probability of the carbon fiber cryosorber, in the 6 to 30 K temperature range [18].

## 4 Cryo-vacuum systems

### 4.1 The LHC arc: a cold bore at 1.9 K

The LHC is a superconducting machine of 26.7 km circumference. The ring is divided in eight 2.8 km long arcs based on regular FODO cells composed of superconducting dipoles and quadrupoles operating at 1.9 K. Between each arc, there are eight room temperature insertions of 575 m long. The two proton beams of 584 mA and 7 TeV collide in four experimental areas located in the middle of the insertions 1,2, 5 and 8.

In the arcs, the vacuum system is made of a cold bore held at 1.9 K surrounded by the superconducting coils assembled in a cold mass. A perforated beam screen operating at 5 to 20 K is inserted into the cold bore to intercept the thermal loads induced by the circulating beam.

During the LHC operation, the proton beams stimulate molecular desorption by photons, electrons and ions bombardment [11, 57, 58, 59]. Consequently, the desorbed molecules shall be pumped to guarantee a beam lifetime of 100 h which is equivalent to a hydrogen pressure of $10^{-8}$ Torr when measured at room temperature.

Fig. 15 shows a cross section of the actively cooled beam screen and the cold bore. The positions of the source of gases are sketched with green arrows. Chemisorbed molecules are desorbed from the beam screen surface by photons, electrons or ions bombardment. These molecules are in turn, either physisorbed onto the beam screen surface, or pumped through the holes where they are condensed onto the cold bore at 1.9 K. Hydrogen, which as a high vapour pressure (see Fig. 8) cannot be physisorbed in large quantities onto the beam screen and is therefore pumped mainly through the holes. Other molecules, which have much lower vapour pressure, are physisorbed onto the beam screen. However, scattered photons, electrons and ions can recycle these molecules into the gas phase. As shown in Section 4.4, when the number of recycled molecules is balanced by the amount of physisorbed molecules, an equilibrium pressure and an equilibrium surface coverage are reached, see Ref. [60].

The equilibrium density, $n_{eq}$, is driven by the pumping speed of the holes, $C$, the photon, electrons and ion fluxes, $\dot{\Gamma}$, and by the primary desorption yields, $\eta$ (see Eq. 20). In consequence, the LHC vacuum system designers have optimised the number of holes and the quality of the technical surface to define the equilibrium gas density at the desired level in order to fulfil the design requirements.

The equilibrium surface coverage on the beam screen, $\theta_{eq}$, is a function of the monolayer coverage, $\theta_m$, driven by the pumping speed of the beam screen, $\sigma S$, the holes pumping speed, $C$, the primary desorption yield, $\eta$ and the recycling desorption $\eta_0'$ at one monolayer, see Eq. (21). The equilibrium surface coverage is therefore less than a monolayer for low sticking probability and large recycling desorption yield.

For illustration of the above effects, studies performed with an LHC mock-up can be used, see Ref. [61]. Fig. 16 shows the hydrogen dynamic pressure measured inside a beam screen when irradiated by synchrotron radiation of 194 eV critical energy at a flux equivalent to one third of the LHC design current [62]. The vapour pressure ($\sim 10^{-10}$ Torr) is subtracted from the data. As seen, in the absence of pumping holes, the dynamic pressure increases owing to the recycling desorption of the physisorbed hydrogen induced by the scattered photons. However, since the experimental apparatus has a limited length of 2 m, the hydrogen pressure level off at $10^{-9}$ Torr due to the external pumping speed located at the extremity. Nevertheless, for much longer lengths as in the LHC arcs, the pressure would have continued to increase well above the LHC design value. Hence, the beam screen shall be perforated to allow the pumping of the recycled hydrogen towards the cold bore. In this case, when perforated with 2 % holes, the hydrogen pressure level off at $4 \cdot 10^{-10}$ Torr, a value well under the LHC design pressure. The experiment demonstrated also that the photon flux may be increased up to a factor 25 while guaranteeing a beam vacuum lifetime of 100 h. This leave much enough margin for an upgrade of the LHC towards HL-LHC (doubling the LHC beam current at 7 TeV) without hindering the beam vacuum lifetime!

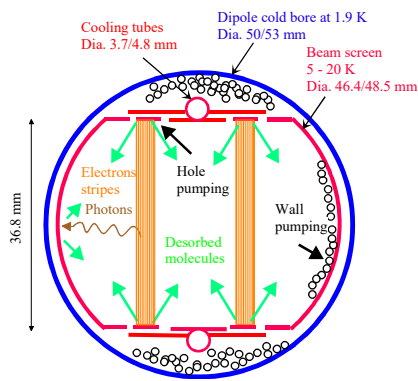
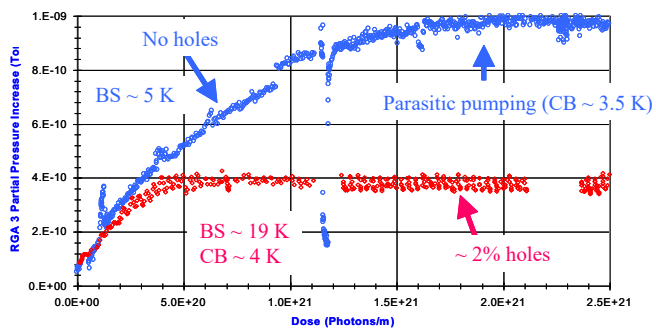

**Fig. 15:** Cross section of the LHC vacuum system

**Fig. 16:** Comparison of the $H_2$ photodesorption between a beam screen with and without holes [62].

### 4.2 The LHC long straight sections: a cold bore at 4.5 K

In the LHC insertions (or long straight sections), the same perforated beam screen technology is used to control the dynamic vacuum. Here, the focusing triplets operate at 1.9 K and the matching section superconducting magnets at 4.5 K. The cumulated length along the ring represents 660 m at 1.9 K and 740 m at 4.5 K. The vacuum system of the focusing triplets operating at 1.9 K behaves similarly to the arc. Yet, the long-term operation of the LHC desorbs several monolayers of hydrogen. Therefore, the saturated vapour pressure (negligible at 1.9 K) will be several orders of magnitude above the design pressure at 4.5 K (see Fig. 7 and Fig. 8). Therefore, as introduced in Sections 3.7 and 3.8, the vacuum system of the magnets operating at 4.5 K requests the use of a cryosorber to be able to adsorb large quantity of gas in order to fulfill the vacuum requirements.

Fig. 17 shows a picture of the LHC beam screen with a carbon woven cryosorber installed on it. An electron shield is clamped on the cooling capillary located in the coaxial space between the beam screen and the cold bore. This shield is installed to protect the cold bore from the heat load dissipated by the electron cloud. The cryosorber, a 5 mm wide black ribbon, is entrenched between the beam screen and the electron shield. By this mean, 200 $cm^2$/m of cryosorber is installed in the narrow space between the beam screen and the cold bore. As said above, the selected cryosorber for the LHC is the woven carbon fiber, see Ref. [55].

To fulfill the LHC performance, the vacuum requirements for the LHC cryosorber are the following:

- Installed outside the beam screen in the shadow of the proton beam
- Operation from 5 to 20 K
- Capacity larger than $10^{18}$ $H_2$/cm$^2$
- Capture coefficient larger than 15 %.

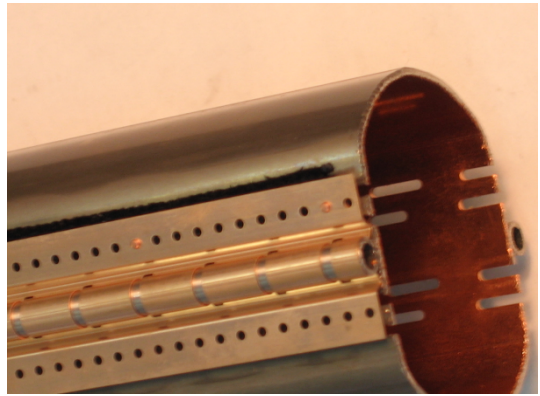

**Fig. 17**: Photograph of an LHC beam screen equipped with a carbon woven cryosorber for use in superconducting magnets operating at 4.5 K.

The operating principle of a cryosorber in a cryogenic beam pipe was validated in a dedicated set up, see Ref. [17]. In this experiment, the cryosorber was activated charcoal. To study the dynamic vacuum, a beam screen was irradiated with synchrotron radiation while the cold bore was held at 70 K. An equivalent of ~ 100 monolayers of $H_2$ was condensed onto the cryosorber before the test. During irradiation, the hydrogen pressure remained below the LHC design pressure while the beam screen temperature was varied in the range 15 to 20 K. Hence, for a gas load twenty times larger than the design capacity, the cryosorber could maintain the stringent vacuum performances.

The LHC cryosorbers installed on the back of the beam screen provide the required $H_2$ capacity and pumping speed up to a certain level of physisorbed hydrogen. When required, the cryosorbers can be regenerated by warming up the beam screen above 80 K while the cold bore is held at more than 20 K in a way to prevent hydrogen adsorption onto it. This action, that requires to empty the cold masses, can take place during machine shutdowns. While the hydrogen is desorbed from the cryosorber, it can be evacuated from the vacuum system by the mobile turbomolecular group. This manipulation is expected to last no more than a few days, see Ref. [63].

### 4.3 Helium leaks in cryogenic beam pipes

The LHC operates with superfluid helium at 1.9 K that has zero viscosity hence, able to flow through any small capillaries. Several thousands of welds have been done to assemble the different components of the LHC during its construction. The bulk materials were carefully produced, selected, and special care was taken to eliminate welds between UHV vessels and liquid helium vessels. Thus, by design, full penetration welds are forbidden along the beam screen cooling capillaries. All the beam screens were leak checked at cryogenic temperature before their insertion into the superconducting magnets. By this approach, the occurrence of a helium leak in the LHC beam vacuum system is greatly minimised.

When a helium leaks appears in a beam tube held at 1.9 K, the gas is physisorbed locally before any possible detection. Consequently, pressure bumps, radiation dose and even magnet quench (a transition from the superconducting to the resistive state of the magnet) may happen during the machine operation. Therefore, the estimation of the pressure evolution due to a helium leak in a superconducting machine is of paramount importance for the cryogenic vacuum designer, see Refs. [64, 65].

At the leak position, a pressure wave develops with time along the beam vacuum chamber, see Ref. [66]. The He condenses locally onto the 1.9 K surface up to a monolayer. Then, at larger surface coverage, the He pressure increases to the saturated vapour pressure (2261 Pa), the He flow moves and the molecules condensed away from the leak. With time, the helium accumulates, and the helium wave can span over several tens of meters without being detected. Consequently, the local pressure bump yields to a local proton loss and a risk of quench.

Fig. 18 schematise the evolution with time of the helium pressure wave along the cold bore axis. The helium leak is located at x = 0. At t = 0, the pressure of the He wave front at the level of the leak, $P_{XF}$ is simply defined by the ratio of the leak rate, $Q$, to the pumping speed of the cold bore aperture, $C_a$. As time goes on, the He wave front progress and saturate the cold bore with helium decreasing the effective pumping speed at the level of the leak. As a result, the pressure at the level of the leak is increasing leading to a linear pressure profile along the cold bore. At t1, the pressure at the level of the leak equals $P_{0,t1}$ and the pressure at the level of the front wave (x = $X_F$(t1)) is $P_{XF}$. At a larger time, t2, the pressure at the level of the leak has increased to $P_{0,t2}$ and the helium wave front is arrived at $X_F$(t2).

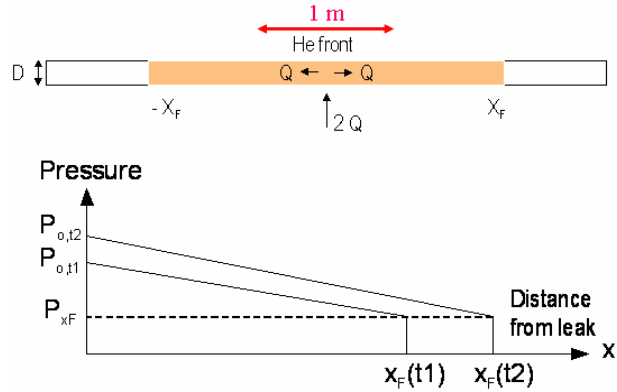

**Fig. 18:** Evolution of the helium pressure with time along the cold bore axis.

A model of the helium propagation wave was developed during the design of the Relativistic Heavy Ions Collider (RHIC) and was used for the LHC [66]. This analytical model predicts the average gas density, $P_{av}$, for a given front length, the He wave front speed, $dx_F/dt$, and the necessary time for the He front to reach a given position, $t_f$.

For a rather good description of the helium front propagation, what matter is the modelling of the steep increase of the He isotherm of Fig. 11. The helium front can be fitted by:

$$\frac{\theta}{\theta_m} = k_1 P^n \qquad (12)$$

Where $\theta/\theta_m$ is the relative surface coverage with $k_1$ and $n$ fitting constants which depends on the temperature. The fitting constants for a stainless-steel tube are: $k_1$ = 1.8, $n$ = 0.0405 at 1.9 K and $k_1$ = 2.47, $n$ = 0.1 at 4.2 K, see Ref. [67].

Equation (15) gives for a leak rate $2Q$, the average He pressure along the He wave of length $2x_F$:

$$P_{av} = \frac{Q\left(1+3\frac{x_F}{8D}\right)}{C_a} \qquad (13)$$

With $D$, the vacuum chamber diameter, $C_a$ = 3.64 $A$ √(T/M), the cold bore aperture conductance, and $A$, the vacuum chamber cross section.

The helium front speed at a point located at $x_f$ is given by Eq. (16). The longer the He pressure wave, the slower the helium front speed is:

$$\frac{dx_F}{dt} = \frac{C_a{}^n Q^{1-n}}{k_1 k_2}\left(1 + 3\frac{x_F}{4D}\right)^{-n} \tag{14}$$

With $k_2 = \pi D\, r_f\, \theta_m\, RT/\mathcal{N}_o$, $r_f$ the internal tube roughness factor, $\mathcal{N}_o = 6.02\ 10^{23}$ mol$^{-1}$, $R = 62.36$ Torr.$\ell$.mol$^{-1}$.K$^{-1}$ or 82.66 mbar.$\ell$.mol$^{-1}$.K$^{-1}$

And the time for the helium front to reach a point located at $x_f$ is given by:

$$t_F = -k_1 k_2 \frac{Q^{n-1}}{C_a{}^n}\left[1 - \left(1 + 3\frac{x_F}{4D}\right)^{n+1}\right] \tag{15}$$

Taking the LHC machine as an example, assuming a helium leak of 3 10$^{-10}$ Torr.$\ell$/s, the He pressure wave after 17 years, will be 212 m long (two LHC cells length: 12 dipoles + 4 quadrupoles!) and slowly progress at a speed of 0.2 cm/h. Detecting such level of leaks with conventional methods will be obviously impossible. However, the proton scattering on the helium may damage the machine or provoked beam dumps.

Conversely, a He leak of 3 10$^{-6}$ Torr.$\ell$/s into the 1.9 K cold bore will provoke a magnet quench within 8 h due to the proton scattering on the He vapour at an average density of 1.8 10$^{11}$ He.cm$^{-3}$. This helium wave will be only 10 m long (~ a quadrupole length) and "quickly" progress at a speed of 57 cm/h.

This He propagation wave model was validated in a dedicated experiment performed in the LHC test string in an LHC type cold bore at 1.9 K and at RHIC during the machine construction period, see Refs. [67, 68].

Fig. 19 shows the evolution of the pressure measured at the level of the leak (an almost flat curve) and the pressure measured 73.5 m downstream to the leak. The predictions of the model are superimposed to the original data. The green curve is the predicted pressure at the level of the leak and the orange curve is the predicted pressure 73.5 m downstream to the leak. The pressure is given at 1.9 K. At t = 0, a leak of 6 10$^{-5}$ Torr.$\ell$/s at room temperature is admitted into the system. It is important to underline that for such a leak rate the pressure at the level of the leak is 10$^{-6}$ Torr corresponding to 10$^{-5}$ Torr at room temperature. Therefore, after less than a few minutes, if a proton beam at 7 TeV would have interacted with such a gas density, the superconducting magnets would have immediately quenched!

However, despite this large leak rate, the helium signal is observed 73.5 m downstream to the leak 20 h after the opening of the leak. For an internal tube roughness factor of 1, the model predicts the apparition of the leak signal after 17 h that is in good agreement with the observations. Obviously, increasing the surface roughness will slow down proportionally the helium front speed.

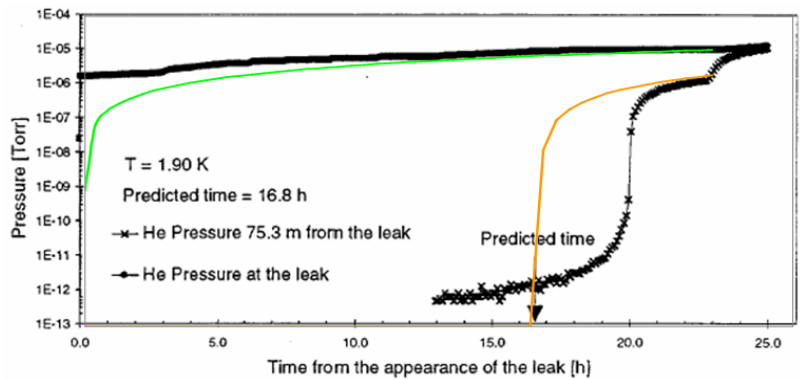

**Fig. 19:** Example of the evolution with time of a He pressure front observed in the LHC string experiment [67]. The predictions of the model are superimposed to the original data.

In LHC, at 7 TeV, $8\ 10^6$ protons/m/s scattered on the cold mass provoke an immediate magnet quench. So, when the amount of proton losses, due to the nuclear scattering on the helium gas in the vicinity of the leak, approaches this number, the quench of the superconducting magnet is triggered. It can be shown that a helium leak rate above $5\ 10^{-7}$ Torr.$\ell$/s shall be detected to avoid the risk of a magnet quench. Lower leak rates can be handled by pumping the beam tube each year but, for instance, a leak rate larger than $10^{-5}$ Torr.$\ell$/s triggers a quench within a day. Therefore, the detection of a potential helium leak is of major importance for the machine operation.

In the arcs, the minimum distance between two consecutive vacuum gauges is 150 m. This distance is by far too large to allow a proper diagnostic of a leak before a quench occurs. Hence, other means of diagnostics are required. The beam loss monitors may be used to detect leak rates below $10^{-6}$ Torr.$\ell$/s. The measurement of the dissipated power in the cold masses seems also another realistic mean of diagnostic. If 1 W/m dissipated in the cold mass can be measured by the cryogenic system, this diagnostic method would allow to detect leak rates up to $10^{-5}$ Torr.$\ell$/s.

In all cases, when a leak is suspected in a given area, an intervention in the tunnel will allow to improve the diagnostic. When possible, the cold mass temperature shall be increased to 4 K to shift the helium front towards the closest short straight section where a vacuum port is located. Afterwards, a residual gas analyzer shall be installed to perform the leak detection. If the leak needs to be monitored during the beam operation, mobile radiation monitors can be installed to follow the radiation wave associated with the pressure wave, see Refs. [69, 70].

When a magnet has already quenched, it can be identified by the triggered diode. During the quench, the cold bore of the magnet is warmed-up to more than 30-40 K. Hence, the helium in the beam tube, if responsible of the quench, is flushed to the nearest unquenched magnet and condensed over a few meters. Again, helium leak detection must be done with a residual gas analyzer.

The repair of a leaking magnet is considered in the following way. For a small helium leak rate, a regular warm up of the cold bore above 4 K and a pump out of the helium every month allows to operate the machine with a reduced beam current. The time estimate to perform such a work is 1 day. For larger leak rates, an exchange of the magnet shall be foreseen. However, before doing such a long and critical work, more time must be invested to be sure that the observed quench is indeed due to a helium leak. For this reason, it is imperative that helium has been clearly identified with a residual gas analyser. Moreover, the position of the leak and its rate shall be known.

For the LHC construction, a few design principles were followed in order to minimise the risk of cold leaks:

- A stringent material selection and specific designs for bellows, cold bores and cooling capillaries compatible with the cryogenic environment were done, see Refs. [71, 72].
- Cold demountable joints and full penetrating welds between beam vacuum and He vessels were forbidden. The cooling capillaries were laser spot welded to the beam screen to avoid full penetration (Fig. 20, left side).
- All beam screens were leaked tested at cryogenic temperature (100 K) and pressurised to 2 bars which results to a detection limit less than $10^{-9}$ mbar.$\ell$/s when operated in the 5-20 K range.
- By design, the welds located at the extremities of the beam screen cooling capillaries are in the insulation vacuum (Fig. 20, right side).

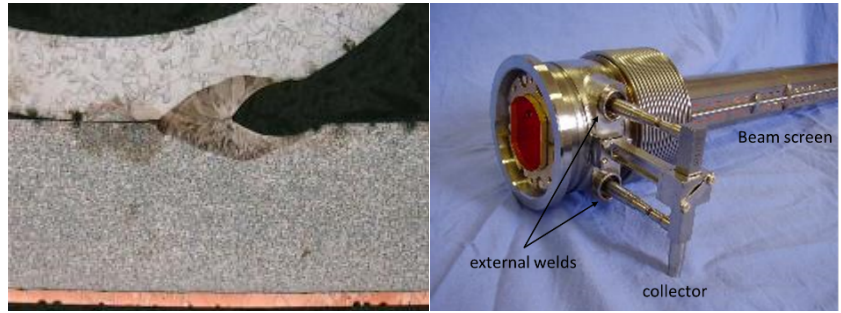

**Fig. 20:** Left, cross section picture of laser welded cooling capillary on the LHC beam screen. Right, picture of an LHC beam screen extremity showing the cooling tubes collector and the external welds located in the insulation vacuum.

A similar design approach was followed for the 4.5 K cryogenic vacuum system of RHIC in order to guarantee a design pressure below $10^{-10}$ Torr along the two rings of 3.8 km long each. Here, sorption pumps based on 300 g of activated charcoal were placed at the magnet interconnects every 30 m to mitigate He leaks and pump $H_2$. Indeed, activated charcoal are routinely used for cryopumps and may be assembled in panels, see Fig. 21. At 20 K, they offer for hydrogen a capacity of ~ $10^{22}$ $H_2$/g (i.e. $10^{21}$ monolayers) and a sticking probability of 30 %.

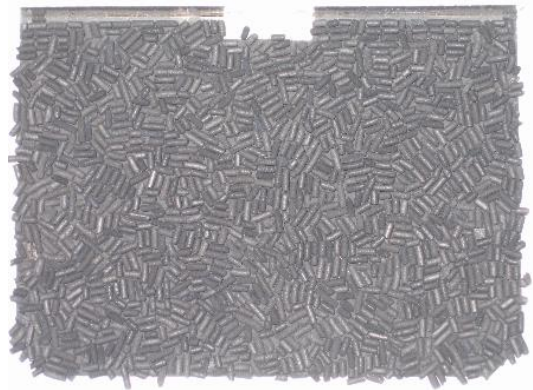

**Fig. 21:** Picture of a 20×10 cm$^2$ activated charcoal cryo-panel.

## 4.4 Design and modelling of a cryogenic beam tube

Modern high energy machines require the use of superconducting technology to keep the tunnel circumference at reasonable dimensions. Table 6 give a summary of the main parameters for the present and future CERN superconducting proton machines. The design, construction and operation of all these cryogenic vacuum systems is (was) really challenging.

The LHC collides proton beams at 14 TeV in their centre of mass. To reach such a high energy, high field superconducting magnets are used to guide the beam along their circular orbit. Synchrotron radiation (SR), emitted in the bending dipole, will stimulate gas desorption under the UV photon irradiation and dissipate heat in the cryogenic environment. The beam screen rationale originates from the prohibitive cryogenic cost of the synchrotron radiation power extraction at 1.9 K. Hence, the LHC beam screen operates at 5-20 K to extract 7 kW of synchrotron radiation power at an electrical cost of 0.6 MW. As shown below (see Fig. 23) and observed with the machine, see Ref. [73], the LHC beam screen transparency set at 4.4 % provide a vacuum beam life time well above the design value.

Today's CERN's flag ship proton storage ring will be shortly upgrade to its high luminosity version, the HL-LHC [74]. The impact of the upgrade on gas desorption and heat loads will push the machine performance towards its limit. This is particularly true for the arc heat load cooling capacity (< 1.35 W/m/ap.) since the sum of synchrotron radiation power and image current heat load distributed around the ring will be multiplied by ~ 2.6 following the beam current increase leaving thereby much less margin for other heat load sources (e.g. electron cloud).

For the future superconducting machines such as HE-LHC and FCC-hh, see Refs. [75, 76], the situation is even much more delicate. As said above, LHC without a beam screen would have required ~ 7 MW of electrical power to maintain the cold bore at 1. 9 K, thus, in the early phases of the design, a beam screen was inserted into the cold bore to intercept the heat load at a higher temperature (5-20 K) in order to intercept the synchrotron radiation heat load at the electrical cost of 0.6 MW thanks to the benefice of the Carnot cycle. However, due to the much larger synchrotron radiation power dissipated in future superconducting machines, the electrical cost will raise to 3 MW for HE-LHC and 75 MW for FCC-hh although a beam screen operating at even higher temperature (45-60 K) is envisaged beneficing even further of a better coefficient of performance of the Carnot cycle. Clearly, these ultimate performances are calling the vacuum designer to develop new concepts & technologies to overcome potential show stoppers that will face the next generation of high energy proton machines (see e.g. Refs. [77, 78]).

**Table 6:** Some parameters of CERN present and future superconducting storage rings.

| Parameter | LHC | HL-LHC | HE-LHC | FCC-hh |
|---|---|---|---|---|
| Beam energy [TeV] | 7 | 7 | 13.5 | 50 |
| Dipole field [T] | 8.33 | 8.33 | 16 | 16 |
| Circumference [km] | 26.7 | 26.7 | 26.7 | 97.75 |
| Beam current [A] | 0.58 | 1.12 | 1.12 | 0.5 |
| Peak luminosity [$10^{34}$ cm$^{-2}$ s$^{-1}$] | 1 | 5 (levelled) | 25 | 5 |
| Bunch population [$10^{11}$] | 1.15 | 2.2 | 1.12 | 0.5 |
| Number of bunch / beam | 2 808 | 2760 | 2 808 | 10 400 |
| Bunch spacing [ns] | 25 | 25 | 25 | 25 |
| Bunch length [ns] | 0.25 | 0.3 | 0.3 | ~ 0.27 |
| SR power [kW] | 7 | 14 | 200 | 4 800 |
| SR power / dipole length | 0.22 | 0.43 | 5.8 | 37.7 |
| SR flux / m [ph/m/s] | 1.0 $10^{17}$ | 2.0 $10^{17}$ | 3.8 $10^{17}$ | 1.7 $10^{17}$ |
| Critical energy [eV] | 44.1 | 44.1 | 314 | 4 300 |
| Image current [W/m/ap.] | 0.14 | 0.52 | 2.3 | 1.7 |
| Electrical power (MW) | 0.6 | 1 | 3 | 75 |
| BS temperature [K] | 5-20 | 5-20 | 45-60 | 45-60 |

In an accelerator, the synchrotron radiation, electron & ion bombardments stimulate the desorption of strongly bounded chemisorbed molecules. The desorption yields range from $10^{-4}$-$10^{-2}$ molecule/photon, $10^{-2}$-1 molecule/electron and 1-10 molecules/ion, see Ref. [59]. As discussed in Section 2, these desorbed species may in turn be cryosorbed on the cold surface but with a much lower binding energy. Hence, the resulting desorption yields are much larger for physisorbed molecules than for chemisorbed ones. At a monolayer, the desorption yields range from $10^{-3}$-1 molecule/photon, 0.1-100 molecules/electron and 1-1000 molecules/ion, see Refs. [79, 80, 81, 82]. In superconducting machines, it is therefore of paramount importance to describe the behaviour of a vacuum system considering the physisorption of molecules.

Since two decades, an analytical model was developed to take into account the observations made with cryosorbing tubes, see Refs. [60, 83, 84, 85, 17]. In this model, the time evolution of the gas density, $n$, and the surface coverage, $\theta$, in a cryogenic vacuum system are described by Eqs. (16) and (17), see Ref. [86].

The gas density increases in the specific beam tube volume, V, is the sum of the 3 gas loads. The first comes from the primary desorption of chemisorbed molecules, $\eta\dot{\Gamma}$, the second, from the recycling of physisorbed molecules in the gas phase, $\eta'\dot{\Gamma}$, and the third, from the vapour pressure, $A\theta/\tau$. The gas load in the volume is reduced by the terms having negative sign that account for the molecular pumping on the cold surface, $\sigma S n$, and the evacuation of the desorbed species outside the beam tube volume thanks to the distributed pumping system, $Cn$. For machines with perforated beam screens inserted in a cold bore, $C$ is defined by the effective pumping speed of the distributed holes. The last term accounts for the molecular axial diffusion along z. In accelerator machines, this term concerns extremity effects such as outgassing originating from a device or external pumping provided by a localised pumping system. In most case, the beam pipe is long enough to neglect the extremity effects which indeed modifies only the gas density locally over lengths smaller than a meter. In practice, the axial diffusion term is therefore negligible for cryogenic pipes with length above a few meters. In particular, the estimation of the average gas density around a ring for the beam vacuum lifetime computation may be done with a good accuracy while neglecting, for instance, the local effect on the gas density provoked by the magnet interconnections.

The increase of the surface coverage of the specific surface, A, is due to the molecular pumping on the cold surface. Conversely, scattered particles recycle the physisorbed molecules in the gas phase, that, together with the vapour pressure effect, lead to the reduction of the surface coverage, hence the negative sign for these two terms in Eq. (17).

$$V \frac{\partial n}{\partial t} = \eta\dot{\Gamma} + \eta'\dot{\Gamma} + \frac{A\theta}{\tau} - \sigma S n - Cn + A_c D \frac{\partial^2 n}{\partial z^2} \qquad (16)$$

$$A \frac{\partial \theta}{\partial t} = \sigma S n - \eta'\dot{\Gamma} - \frac{A\theta}{\tau} \qquad (17)$$

With:

- $V$ the beam tube volume per unit axial length ($\pi a^2$ for a circular tube of radius, $r$) [m$^3$.m$^{-1}$]
- $A$ the vacuum chamber surface per unit axial length [m$^3$.m$^{-1}$]
- $A_c$ the tube cross section area ($\pi a^2$ for a circular) [m$^2$]
- $D$ the Knudsen diffusion coefficient ($2/3\ r\ \bar{v}$ for a circular tube with mean molecular speed, $\bar{v}$) [m$^2$.s$^{-1}$]
- $\eta$ the primary desorption yield [molecules.particle$^{-1}$]
- $\eta'$ the recycling desorption yield [moleculesparticle$^{-1}$]
- $\dot{\Gamma}$ the flux of desorbing particles [particles.m$^{-1}$.s$^{-1}$]
- $\tau$ the sojourn time of physisorbed molecule [s$^{-1}$]
- $\sigma$ the sticking probability
- $S$ the ideal pumping speed [m$^3$.s$^{-1}$.m$^{-1}$]
- $C$ the distributed pumping speed [m$^3$.s$^{-1}$.m$^{-1}$].

It is instructive to study some specific solutions of the above equations. Let's consider a long cryogenically cooled beam tube without distributed pumping. In the quasi-static case (for "slow" evolution of the gas density with time i.e. d$n$/d$t \sim 0$), the time solutions to the above set of equation with $A_c D = 0$ and $C = 0$ are:

$$n(t) = \frac{\eta\dot{\Gamma}}{\sigma S} + \frac{\eta'\dot{\Gamma}}{\sigma S} + \frac{1}{\sigma S}\frac{A\theta}{\tau} \qquad (18)$$

$$\theta(t) = \frac{1}{A}\int_0^{\Gamma} \eta \, d\Gamma \qquad (19)$$

In a long cryogenically cooled beam tube, the gas density is the sum of the primary desorption, recycling desorption and vapour pressure. The surface coverage originates from the continuous desorption of chemisorbed molecules which ultimately are physisorbed on the surface owing to the closed geometry.

Fig. 22 shows the evolution with accumulated photon dose of the hydrogen density and relative surface coverage for a hypothetical non-perforated LHC beam screen. In this case, the 100 h lifetime limit is reached after only a 3 hours of operation with nominal current. The first period is dominated by the primary desorption, then by the recycling desorption. When the relative surface coverage approach one monolayer, the beam tube gas density becomes dominated by the vapour pressure and reach the saturated gas density after a few ten hours of operation. The parameters used for this computation are $\dot{\Gamma} = 10^{17}$ ph.m$^{-1}$.s$^{-1}$, $\eta = 10^{-4}$ H$_2$/ph, $\eta'_0 = 1$ H$_2$/ph. Here, we have assumed that the recycling desorption yield of hydrogen scales linearly with the surface coverage like $\eta'(\theta) = \eta'_0 \, \theta/\theta_m$ where $\theta_m$ is the monolayer surface coverage. The sticking probability is assumed to vary linearly with relative surface coverage starting at 0.2 and reaching 1 at a monolayer (see Fig. 4).

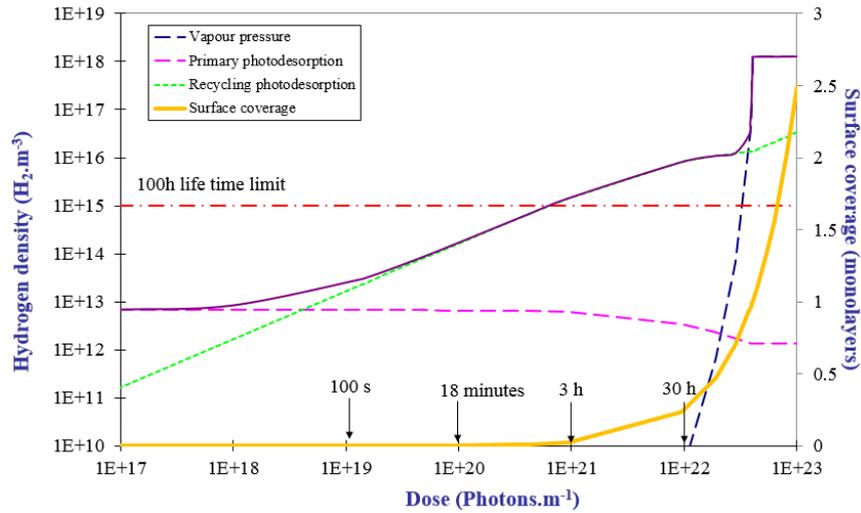

**Fig. 22**: Evolution of the hydrogen gas density and relative surface coverage for a non-perforated LHC beam screen.

Let's consider now a long-perforated beam screen ($A_cD = 0$, $C \neq 0$), that allows for the distributed pumping of gas along the beam tube volume. In the quasi-static case ("slow" evolution with time of the gas density and surface coverage i.e. d$n$/dt ~ 0 and d$\theta$/dt ~ 0), the gas density and surface coverage reach an equilibrium given by the solution of Eqs. (16) and (17):

$$n_{eq} = \frac{\eta \dot{\Gamma}}{C} \qquad (20)$$

$$\Theta_{eq} = \left(\frac{\sigma S}{C} \frac{\eta}{\eta'_0}\right) \theta_m \qquad (21)$$

The equilibrium gas density is simply given by the most basic formula in vacuum science and technology i.e. the ratio of the gas load over the pumping speed of the beam screen perforations! The higher the perforation's conductance, the lower the equilibrium gas density and surface coverage are. Reducing the gas density in a superconducting machine in order to increase the beam vacuum lifetime is therefore straightforward. A pre-treatment (e.g. ex-situ baking, ex-situ ion or electron conditioning etc.) of the vacuum chamber wall may be used to reduce the primary desorption yield and / or an increase of the beam screen's perforation size and number are the solutions to properly design your cryogenic vacuum system.

To be noted also is that the level of the recycling desorption yield does not modify the equilibrium density, however, a large recycling desorption yield reduce the equilibrium coverage due to the flushing of the gas from the beam screen towards the beam screen's perforations. Taking the LHC beam screen subjected to synchrotron radiation irradiation as an example, with 4.4 % transparency and a Clausing factor of 0,5; the relative equilibrium coverage of hydrogen is in the range from $5\times10^{-4}$ - $5\times10^{-3}$ whereas it range from 0.5-5 for the other gases due to their much lower recycling yield [79]. Hence, one may accumulate a few monolayers of molecules on the beam screen except for $H_2$. However, in the presence of electron cloud and the continuous bombardment of the beam screen by the electrons, the relative equilibrium for all molecular species is greatly reduce, well below a monolayer, due to the large recycling yield for electrons [80]. Therefore, no molecular species are expected to be cryosorbed on the inner side of the LHC (or HL-LHC) during its operation.

Fig. 23 shows the hydrogen density and relative surface coverage evolution during operation of the LHC as a function of the accumulated photon dose. After a couple of minutes, the equilibrium density is reached. At this moment, the relative surface coverage ceases to increase: the amount of gas physisorbed on the surface is exactly balanced by the amount of gas recycled in the gas phase. Everything is happening as if all the primary desorbed gas is pumped by the beam screen perforation, hence the equilibrium gas density. The input parameters used in this computation were the same as the one used in Fig. 22. At equilibrium, the gas density is set at $6\ 10^{13}$ $H_2/cm^3$ for a relative surface coverage of $8\ 10^{-4}$ monolayers ($2.5\ 10^{12}$ $H_2/cm^2$). Assuming the sole presence of $H_2$ in the beam tube, the corresponding beam vacuum lifetime is ~ 1 650 h, well above the LHC specifications!

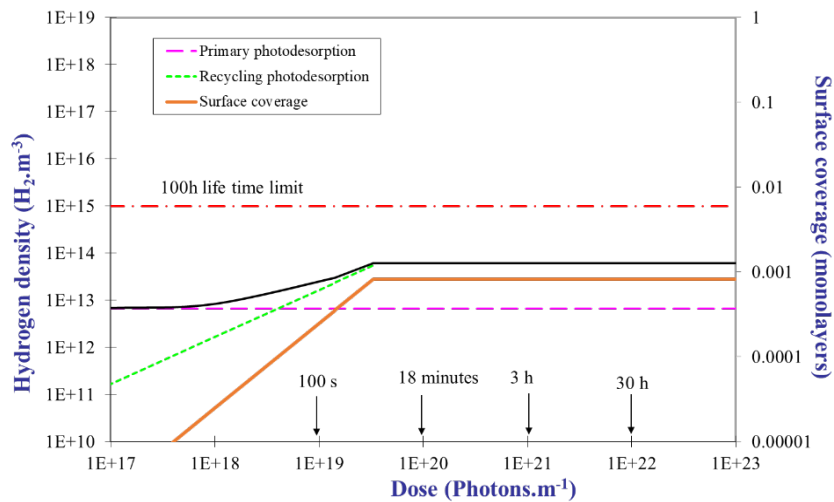

**Fig. 23:** Evolution of the hydrogen gas density and relative surface coverage for an LHC beam screen with 4.4% transparency.

### 4.5 Operation with cryo-vacuum machines, the examples of LHC, SIS100, RHIC, ISR, LEP2 and HIE-Isolde

Since several decades, more and more vacuum systems of accelerator machines operate at cryogenic temperature. A few examples taken from the operation of these machines are discussed below to illustrate the importance of the cryogenic vacuum science and technology in everyday work.

LHC is presently the most sensitive machine to cryo-vacuum aspects. Indeed, before or during beam operation, some gas may be cryosorbed on the beam screen. As a result, the beam-vacuum interactions may be modified to provoke: background to the experiments following vacuum transient, particle loss triggering beam loss and beam dump, unwanted heat load on the cryogenic system etc.

As shown in Eq. (20) and (21), when subjected to particle irradiation, the perforated beam screen provides a gas density and surface coverage equilibrium. Hence, any excessive gas coverage on the beam screen will result in a vacuum transient towards the equilibrium value. In Fig. 24, the upper curve shows a pressure transient provoked by synchrotron radiation irradiation of a monolayer of hydrogen pre-condensed on the beam screen surface [63]. Under the hydrogen recycling, the pressure increases till $3 \cdot 10^{-8}$ Torr and decrease, in a couple of hours, towards the equilibrium pressure ($3 \cdot 10^{-10}$ Torr) defined by the holes pumping speed. The gas & surface balance set of equations were used to fit the parameters shown on the plot. The lower curve in the figure shows the natural vacuum transient in the absence of gas cryosorbed on the surface. Therefore, the gas coverage on a cold surface must be minimised to avoid undesirable & large pressure transients.

The cryosorption of gas may also result in large heat load caused by the beam induced electron cloud phenomena [87, 88]. Indeed, the secondary electron yield of cryosorbed gas can be well above one [89, 90] leading in some circumstances to strong multipacting, thereby, large heat load in a cryogenic vacuum system. Fig. 25 shows the pressure and heat load evolution in an LHC like vacuum system installed in the CERN SPS machine. From day 0 to 4, the pressure inside the beam screen decrease by a factor 20 whereas the heat load dissipated into the cryogenic vacuum system increases till 8 W/m when only 2×72 bunches with $10^{11}$ proton/bunch circulates. At day 5, a warm-up at 240 K, to desorb $H_2O$ from the beam screen, results, from day 6, in strong a reduction of the dissipated heat load. Again, minimising the beam screen's surface coverage is mandatory to avoid unwanted heat loads on the cryogenic system.

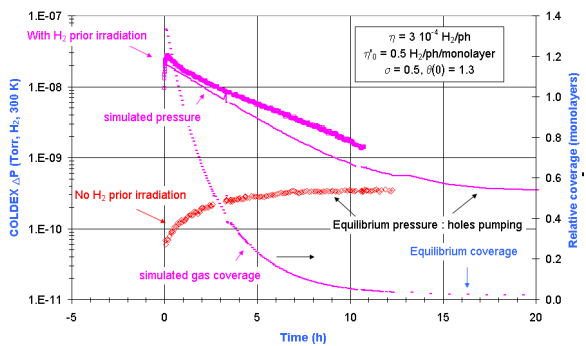 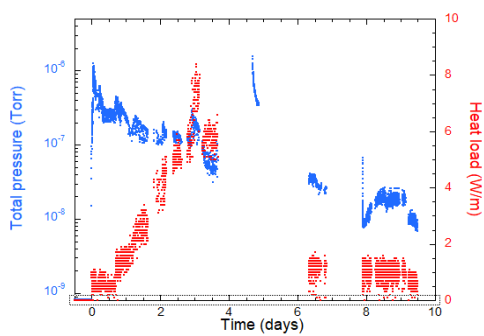

**Fig. 24:** Hydrogen vacuum transients provoked by synchrotron radiation [63].

**Fig. 25:** Pressure and heat load evolution provoked by beam induced multipacting [87].

For these reasons, in LHC, a specific procedure is applied to minimise the amount of cryosorbed gas on the beam screen, see Ref. [91]. Before the cool down of a cryogenic beam pipe, a room temperature pump down for at least 5 weeks using a mobile turbomolecular pumping system allows to evacuate as much as possible $H_2O$ from the vacuum vessel. During the cool down process, the cold bore is cooled down first without active cooling of the beam screen to let a temperature difference between the cold bore and beam screen in order to maximise the chances of cryo-condensation on the cold bore. Then, the beam screen temperature may be stopped at ~ 90 K for a couple of hours until the cold bore reaches less than 20 K. Finally, active cooling of the beam screen is triggered to reach the final operating temperatures: 1.9 K for the cold bore and 5-20 K for the beam screen.

If an excess of cryosorbed gas on the beam screen is suspected/present (e.g. following a mis-controlled cool-down or a magnet quench), heaters can be used to warm up the beam screen to ~ 90K in a way to flush the gas towards the 1.9 K cold bore. This method allows to partially regenerate a cryogenic surface by evacuating most of the common gas except $H_2O$ mitigating the impact of cryosorbed gas on the LHC operation, see Refs. [63, 91].

The next generation of cryogenic machines may also potentially suffer of the negative impact of cryosorbed case on the cryogenic beam tube. This is the case of SIS-100, a synchrotron presently under construction for the Facility for Antiproton and Ion Research (FAIR) at Darmstadt. This 1083.6 m long machine will accelerate and deliver radioactive beam pulses up to 35 GeV/u. It has 895 m of its length operating in the 5-10 K range at a design gas density less than $8 \cdot 10^{11}$ $H_2/m^3$. During operation, the trajectory of the circulating ions may be modified due to charge exchange thereby stimulating strong molecular desorption when bombarding the vacuum chamber walls. For this reason, collimators are placed at specific location around the ring to control and minimise the gas load. To provide the desire beam characteristics, the superconducting dipole field is ramped with 4 T/s at 1 Hz. Hence, the induced Eddy currents provoke a non-uniform and non-constant heat load across the beam pipe. Consequently, the temperature along the vacuum chamber cross section varies from 5 to 15 K potentially stimulating vacuum transients for large enough surface coverage. Solving Eqs. (16) and (17) in a time dependant 1D model of the machine, it has been shown that after ~ 180 days of operation, the hydrogen gas density limit will be reached. Therefore, a regular warm up of the cryogenic beam pipe to 20 K is envisaged to evacuate the cryosorbed gas towards an external pumping system, see Refs. [92, 93].

Machines contemporary to LHC are also sensitive to cryosorbed gas. Indeed, during the years 2004-2005, large pressure rises (up to $10^{-7}$ mbar) were observed in the arcs and triplets of RHIC. The pressure rises led to strong proton losses caused by the residual gas pressure increase stimulated by ion and/or electron bombardment of the 4.5 K cold bore. To solve the issue, a dedicated pump down stage was implemented in the vacuum vessel evacuation procedure. A pressure of ~ $10^{-6}$ mbar before cool-down was obtained in a way to minimise the amount of cryosorbed gas on the beam pipe, see Ref. [94].

In fact, the impact of cryosorbed gas on machine design and operation has been known since several decades. Indeed, at the CERN Intersecting Storage Rings, two cold bore experiments were conducted from 1975-1980 in order to prepare the future use of superconducting magnets and study the vacuum stability and the condensation of gas [95]. Among other results, pressure spikes up to $10^{-8}$ mbar were recorded with 30 A proton beams when ~ 100 monolayers of $N_2$ were condensed on the surface to simulate an air leak. The suspected origin of the spike was a breakdown in the nitrogen film which led to gas desorption. Air leaks in cold systems are therefore not acceptable. That was recently recalled for the LHC with the so-called 16L2 incident that provoked several beam dumps due to unwanted air condensation at a magnet interconnect [96]. Fortunately, a (partial) warm-up of the system allowed to solve the issues.

The impact of cryosorbed gas on machine performance is also observed in superconducting RF technology. At CERN, the LEP2 machine brought into collision electrons & positrons at ~ 190 GeV [97]. To overcome the beam loses by synchrotron radiation, the conventional normal conducting cavities were replaced from 1996 onwards by 6 MV/m, 352.2 MHz, Nb coated, superconducting RF cavities operating at 4.5 K. During the construction phase of the superconducting RF cavities, a deconditioning of the coupler surface was noticed. The origin was traced back to water condensation on the outer surface of the coupler that produced multipacting at cryogenic temperature stimulating further water desorption by RF heating. The issue was solved by 1) an in-situ bakeout at 200°C of the RF coupler before cool-down and, 2) a 2.5 kV negative bias applied on the centre conductor to modify the electron kinetics to mitigate multipacting [98]. Similar observation was recently done during the construction of the CERN High Intensity and Energy Isolde (HIE-Isolde). The project allows for radioactive ion beams acceleration, from mass 6 to 224, at 10 MeV/u. [99]. For this purpose, 100 MHz superconducting quarter-waves Nb coated cavities are used. The conditioning process of the cavity is performed at 200 K before cooling down to 4.2 K for operation. If multipacting occurs after cool-down, a recipe consists in warming up the cavity to 20-30 K to redistribute and pump away the cryosorbed gas [100]. In conclusion, in superconducting technology also, the amount of cryosorbed gas on a cold surface shall be minimised to guarantee a proper operation of the device when submitted to extreme conditions.

# 5 Summary

The knowledge of Vacuum Science & Technology at cryogenic temperature is a key ingredient for the design, operation and understanding of present and future accelerator machines. The present lecture discussed the main concepts, elements, data and models required to master this discipline.

The study of the interactions of a molecule with a surface held at cryogenic temperature results in the following outcome:

- Depending on the binding energy, molecules can be physisorbed for very long period on cryogenic surface;
- The sticking coefficient characterises the pumping speed of a cryogenic surface whereas the capture coefficient the pumping speed of a device;
- At cryogenic temperature, thermal transpiration correction shall be applied;
- The vapour pressure is the equilibrium pressure as a function of gas coverage on a cryogenic surface;
- When many molecules are condensed on a cryogenic surface (many monolayers), the saturated vapour pressure follows the Clausius Clapeyron law;
- Adsorption isotherms vary very much with the conditions and the nature of the surfaces, some analytical models are available in the literature;
- Porous materials (cryosorbers) can adsorb many monolayers of gas without reaching the saturated vapour pressure, they can be characterised by the roughness factor;
- The design of a cryogenic beam tube vacuum depends on the operating temperature and the environment. In the LHC, perforated beam screens provide the required vacuum level and cryosorbers allows a cold bore operation above 3 K;
- He leaks in a cryogenic beam tube can be modelled but their detection is cumbersome except by the beam itself! A specific engineering approach is required to avoid cold leaks;
- The particle (electron, photon, ions) bombardment inside a cryogenic beam tube and the resulting pressure (gas density) and surface coverage can be analytically modelled.
- Molecular physisorption and condensation can be strongly detrimental for the operation of vacuum system held at cryogenic temperature. For this reason, appropriate design, surface treatments and minimisation of the sources of gas is required.

The ingredients developed in the above lecture are useful for the present or future vacuum scientist/engineer in charge of the design or operation of an accelerator operating at cryogenic temperature.

**Epilogue**

Fig. 1 is a picture taken in August 2009 during the LHC repair period (following the sector 3-4 accident), a few months before the first beam turns on the 20$^{th}$ of November 2009.

The time for warming up and cool down a complete LHC arc last several months. It is also a costly and risky exercise. The 11$^{th}$ July 2009, a large helium leak (0.3 mbar.$\ell$/s) appeared in an insulation vacuum located at the right extremity of the sector 2-3. For time and cost reasons, a partial warm up of the arc was decided to perform the repair of the faulty flexible located in the insulation vacuum sector A7L3.M (that spans from quadrupoles Q7L3 to Q11L3 including the distribution feed box, DFBA). To do so, it was decided to stop the active cooling of the arc magnets chain, vent and open to atmosphere the insulation vacuum subsector to allow access for repair. Obviously, in the meantime, all the necessary precautions for personnel / material safety and against water condensation in the machine were taken. Unfortunately, the LHC arc vacuum system suffer from an important non-conformity which implies a systematic inspection of any RF bridge that has warmed-up above ~100K and risk to buckle.

Therefore, a visual inspection of the plug-in-modules, QBQI.12L2, located between magnets A12L3 and Q11L3 was needed.

Being non-condensable above ~ 20 K, the beam vacuum was vented with neon using ~ 40 bar.m$^3$ per beam line at 40 K (opposed to 6 bar.m$^3$ per beam line when vented at room temperature). Then, the magnets in the insulation vacuum sector A7L3.M (~200 m long) were warmed up to 300 K. The remaining part of the arc (~ 2.6 km) was let at ~ 40 K, drifting slowly to higher temperature at a rate of 1.1 K/day, leaving at least a month for a safe repair without the risk to damage the RF fingers. As a precaution, the flow of neon into the beam tube was maintained during the whole repair process to minimise the air back streaming into the pipe. In order to inspect the potentially faulty RF bridges in QBQI.12L3, both plug-in-modules at QQBI.9L3 were cut to allow the insertion of an endoscope about 100 m away from the RF bridges.

The picture shows the inside of the LHC beam screen in the magnet A12L3 when held at 60 K. The white stain is the condensation of impurities ($O_2$, $H_2O$, $N_2$) present in the Ne bottle (purity level of 99.999 %). The estimated quantity is 250 Torr.$\ell$, which translates into about $5 \cdot 10^{18}$ molecules/cm$^2$ (5 000 monolayers!) condensed at the beam screen extremity over about a meter. A neon's impurity cleaning using a NEG cartridge would have provided an ultrapure injection of gas in the cold system.

Under vacuum, the corresponding saturated vapour pressure as a function of temperature for these gases would have been negligible at the nominal operating temperature of the beam screen, below 20 K (see Fig. 7). However, the circulating proton beams may have strongly interacted with this condensed layer provoking particle losses (hence beam dumps), vacuum transients (see Fig. 24), heat loads (see Fig. 25), etc.

Fortunately, during RUN 1, the LHC operated at 4 TeV with 50 ns bunch spacing and a maximum beam current of 420 mA. This set of parameters results in moderated synchrotron radiation ($4 \cdot 10^{16}$ photons/m/s, 8 eV critical energy) and electron cloud (heat load < 0.2 W/m). This conjugated with the beam screens warm up above 100 K during the winter technical stops, implied that the dynamic effects associated with the gas condensation were not observed. Finally, during the Long Shutdown 1 (2013-2014), the warming up until room temperature of the complete LHC ring for the repair campaign of the magnet interconnections, allowed to pump away these condensed gasses, securing a proper operation of the machine for the next RUNs.

**Acknowledgements**

It is my pleasure to acknowledge the support and expertise of my colleagues from the CERN Vacuum Group and many collaborators across the world without whom this lecture would ever had existed. I would like also to warmly thanks the organisers and the successive directors of the CERN Accelerator Schools for their regular invitations to give lectures.

**Some books addressing the cryogenic vacuum science & technology**
- The physical basis of ultrahigh vacuum. P.A. Redhead, J.P. Hobson, E. V. Kornelsen. Chapman and Hall, London, 1968.
- Scientific foundations of vacuum techniques. A. Dushman, J.M. Laferty. John Wiley & sons, New-York, 1962.
- Physical adsorption of gases. D.M. Young, A.D. Crowell. Butterworths, London, 1962.
- Cryopumping- Theory and Practice. R.A. Haeffer. Oxford science publications, Oxford, 1989.
- Vacuum technology. A. Roth. North-Holand, Elsevier, Amsterdam, 1990.
- Capture pumping technology. K.M. Welch. North-Holand, Elsevier, Amsterdam, 2001.
- Vacuum in particle accelerators: modelling, design and operation of beam vacuum systems. O. B. Malyshev. Wiley VCH, Weinheim, 2020.